\documentstyle[preprint,aps,epsf,psfig]{revtex}
\tightenlines

\begin{document}
\draft

\title{Thermal Resonance in Signal Transmission}

\author{Ramon Reigada\footnote{Permanent address:
Departament de Quimica-Fisica, Universitat de Barcelona,
Avda. Diagonal 647, 08028 Barcelona, Spain},
Antonio Sarmiento\footnote{Permanent address:
Instituto de Matematicas, UNAM, Avenida Universitaria s/n,
Chamilpa, Morelos 62200, M\'{e}xico}
and Katja Lindenberg}

\address
{Department of Chemistry and Biochemistry 0340\\
University of California San Diego\\
La Jolla, California 92093-0340}

\date{\today}
\maketitle

\begin{abstract}
We use temperature tuning to control signal propagation in simple
one-dimensional arrays of masses connected by hard anharmonic
springs and with no local potentials. In our
numerical model a sustained signal is applied at one site
of a chain immersed in a thermal environment and
the signal-to-noise ratio is measured at
each oscillator. We show that raising the temperature can lead to enhanced
signal propagation along the chain, resulting in thermal resonance effects
akin to the resonance observed in arrays of bistable systems.
\end{abstract}


\section{Introduction}
\label{intro}

In the past few years it has become abundantly clear that the
presence of noise in nonlinear systems may lead to an
enhancement of a number of often desirable features such as energy
localization and mobility and the detection and propagation of weak
signals. The interplay of stochasticity and nonlinearity that
amplifies the system response is a cooperative phenomenon whose
detailed nature depends on the particular structure of the system
and the forces acting upon it\cite{first,rev}.  One manifestation of the
interplay is found in the phenomenon called stochastic resonance, which has
been invoked in a wide range of
physical~\cite{bistable,physical1,physical2},
chemical~\cite{Reichl,exp,jung,chem}, geological~\cite{first,geo},
and biological~\cite{jung,bio} systems.
Recent literature, including our
own work,~\cite{our1,our2,our3} has focused on spatially
extended systems~\cite{ojalvo} including
noise-enhanced propagation in coupled arrays of bistable
units~\cite{bistable,ditto},
excitable media~\cite{exp,jung,jung2},
reaction-diffusion systems~\cite{wio}, and dynamics and signal propagation
in cardiac tissue~\cite{keener,cole}. It has been repeatedly noted
that {\em discrete} extended systems pose particular mathematical
challenges that have barely been explored in spite of the fact
that many physical systems are intrinsically
discrete~\cite{Willis,Laroche,Campbell,Martinez,Giardina}.

The ubiquitous picture of stochastic resonance involves a particle
moving in a double-well potential subject to a weak external
signal that periodically changes the potential by alternately
raising and lowering the wells \cite{rev}. The signal is ``weak"
if the periodic force is too small to cause the particle to scale
the barrier between the wells. Nevertheless, an appropriate random
force is sufficient to cause the particle to cross over the
barrier even in the absence of a deterministic signal. In the
simultaneous presence of a weak signal and a sufficiently weak
noise, the transitions over the barrier occur
rarely and at a rate determined by the noise intensity.  These
transitions are slow compared to
the frequency of the deterministic
signal; the transition rate then carries little information about
the signal. At the other extreme, when the noise is strong it induces
rapid transitions that are again essentially uninfluenced by the
frequency of the signal. At an optimal noise intensity, however,
the mean first
passage time associated with the noise and the frequency of the
signal are in synchrony (stochastic resonance), and the passage
from one well to the other carries maximal information about the
signal frequency.

A less ubiquitous but nonetheless important occurrence of
stochastic resonance (that has been called ``nonconventional" by its 
discoverers) arises for particles moving in nonlinear
{\em monostable} potentials~\cite{Dykman,Stocks}.  It is argued that
stochastic resonance can be expected to occur in any single-well
underdamped system for which
the spectral density of the fluctuations of the system in the absence
of a periodic signal exhibits a well-resolved narrow peak that grows
faster than quadratically with temperature.  The effect is confirmed
via analog simulations of a single-well Duffing
oscillator~\cite{Dykman,Stocks} and of a SQUID
loop~\cite{Kaufman}.  More recently, stochastic resonance at higher
harmonics in
monostable systems was ascertained for an overdamped system when the
nonlinearity is not concentrated at the equilibrium
position~\cite{Grigorenko}.

Recent developments in the field have
generalized these ideas to linearly coupled arrays of bistable
oscillators~\cite{bistable,physical1,ditto}. A signal with the help of the
noise in these arrays can cause a ``phase jump." If the noise is
sufficiently weak, the phase
jump travels in the form of a moving kink (strong noise causes random phase
jumps that make it difficult to separately identify a phase jump
associated with the
signal).    The creation or destruction of a phase kink is an
activated process, i.e., the
signal and/or noise must be sufficiently strong to cause a transition
from one well of one of the bistable potentials to the other.  The presence
of such kinks is associated with an ``energy gap:"  it takes a
finite amount of energy to destroy a kink.  The language used in this
description was originally borrowed from the kink soliton context.

Reported instances of stochastic enhancement and stochastic
resonance in {\em extended} arrays involve coupled (overdamped)
{\em bistable} units.  Herein we show that enhanced propagation
can be achieved through thermal tuning
of even simpler discrete arrays of masses connected by {\em monostable}
anharmonic springs (with no local potentials).
The signal is identified with an amplitude that
exceeds (by a predetermined amount) that due to the thermal
background.  Here there is no activation process and no energy gap,
and the signal can simply disperse or dissipate.
The language appropriate to this case is akin to that originally
associated with envelope solitons.  We focus on the
propagation distance and amplitude along the chain of a signal
continuously applied at one site of a 1D array,
and compare results for harmonic and hard anharmonic chains.
In particular, we show that when the anharmonic chain is immersed
in a thermal bath, it is possible to maximize the distance of
propagation and the amplitude of the signal at a given site by
tuning the temperature to particular optimal values. We call this
phenomenon {\em thermal resonance}.  Our systems are in
general not overdamped and
thus include inertial contributions to the motions of the
masses. Noise and damping represent a realistic thermal
environment with a tunable temperature and dissipation that obey an
appropriate fluctuation-dissipation relation. We have found and
reported elsewhere~\cite{our2} that
the propagation of an energy pulse in a hard anharmonic array
can be enhanced by immersion in a thermal bath, and
that hard anharmonicity in the springs causes a tight and
persistent packing of the energy. Those results
suggest the possibility of a thermal resonance in the
transport of a sustained external signal in these simple arrays.

In Section~\ref{model} we present our model and some details of
the numerical integration of the equations of
motion.  Our characterization of a thermal resonance is
presented in Section~\ref{character}, and our main results are
shown in Section~\ref{results}.  In
Section~\ref{parameters} we discuss the dependence of our results
on different parameter models.  Section~\ref{conclusion} contains
our summary and conclusions.

\section{The Model and Numerical Procedure}
\label{model}

Our model consists of a
one-dimensional chain of $N$ unit-mass sites, each connected to its
nearest neighbors by either harmonic (quadratic) or hard
anharmonic (quartic) springs.  Accordingly, the Hamiltonian of the
array is
\begin{equation}
H = \sum_{n=0}^{N-1}\left(\frac{p_n^2}{2} + \frac{k}{2}
(x_n-x_{n-1})^2
+\frac{k'}{4}(x_n-x_{n-1})^4\right)~,
\label{hamiltonianr}
\end{equation}
where $k$ and $k'$ are the harmonic and anharmonic coupling
constants respectively. Thermalization of the
chains is achieved through a Langevin prescription for coupling
the
system to a heat bath. The stochastic equations of motion for
sites $n=1,\cdots,N-1$ are obtained from the Hamiltonian augmented
by the usual Langevin forces:
\begin{eqnarray}
\ddot x_n  & = &  k(x_{n-1} - x_n) - k(x_n - x_{n+1})  +  k'(x_{n-1} -
x_n)^3\nonumber\\ [12pt]
&& - \ k'(x_n - x_{n+1})^3 -  \gamma \dot x_n +f_n(t)~,
\label{lang}
\end{eqnarray}
where a dot represents a derivative with respect to time. The $f_n(t)$ are
zero-centered, Gaussian, $\delta$-correlated fluctuations that satisfy the
fluctuation-dissipation relation at temperature $T$,
\begin{equation}
\langle f_n(t) f_{n'}(t') \rangle = 2
\gamma k_B T \delta_{n,n'} \delta(t-t')
\label{fdr}
\end{equation}
($k_B$ is Boltzmann's constant).  We impose periodic boundary conditions,
so that $x_N\equiv x_0$.
A sustained signal is applied to the site $n=0$
that determines its velocity at all times:
\begin{equation}
\dot x_0  = A  \sin ( \omega_0 t)~.
\end{equation}
The positions and momenta of all the other sites are
otherwise ``free'' and determined by the equations of motion. We study the
propagation of this signal along the chain as a function of the temperature.
This particular way of applying a signal is of course not unique (e.g.
one might apply an oscillatory {\em force} instead), but we
have ascertained that the results are insensitive to the detailed choice.

An analytic solution of this problem is not available for an
anharmonic chain, so
we must rely on numerical integration, which is performed using
the second order Heun's method (equivalent to a second order
Runge-Kutta integration)~\cite{Gard,Toral}.  The
time step is determined by the period of oscillation of the
velocity of the first site, $\tau = 2\pi/\omega_0$, through the relation
$\Delta t = \tau/2^{12}$.
For each simulation, the system is
initially allowed to relax to thermal equilibrium.
For all the simulations presented, this is achieved in less
than $20$ units of the dimensionless time.
Typically, after a transient that
is longer the further the site is from site 0 (and thus a measure of the
velocity of propagation), each site settles into stationary behavior
that is a
mixture of thermal motion and response to the signal. Sites that are far
from the signal never exhibit this transient (thus indicating a finite
distance of propagation) and simply continue their thermal motion.
At any given site that {\em is} reached by the
signal, one can observe the amplitude of the motion associated
with the signal over and above the thermal motion.

Our interest here lies in demonstrating resonances in the dependence on
temperature
of the propagation distance and velocity and of the response amplitudes
once stationarity has been achieved (in all our
simulations, each site has settled into its long-time behavior after 100
first-site oscillations). Possible energy return effects around the
periodic chain are prevented by making the chain sufficiently long
and/or sufficiently increasing the dissipation parameter of distant sites.
Our chains typically consist of 70 sites with a large dissipation
at sites 27--32 (these numbers can easily be varied).
Our ``measurements"
are then taken over 80 oscillation periods and
non-zero-temperature results are averaged over 500 realizations. All of
these choices (equilibration time before applying the signal, integration time
step, transients, length of chain, and number of realizations) have been
carefully tested.

\section{Characterization of Thermal Resonance}
\label{character}
We must choose a sensible response variable to characterize the
behavior of our array.  When stochastic resonance is studied in
arrays of bistable potentials, the system response is usually
analyzed in terms of a crossover time series that characterizes
the transitions of each bistable element from one well to the
other.  In excitable media a reasonable response variable involves
firing times of the individual elements.  In our system the most
convenient choice is the velocity of each
site because it corresponds directly to the applied signal, and
because the time average of the velocity in the stationary state
vanishes at any temperature.

\begin{figure}[htb]
\begin{center}
\leavevmode
\epsfxsize = 3.0in
\epsffile{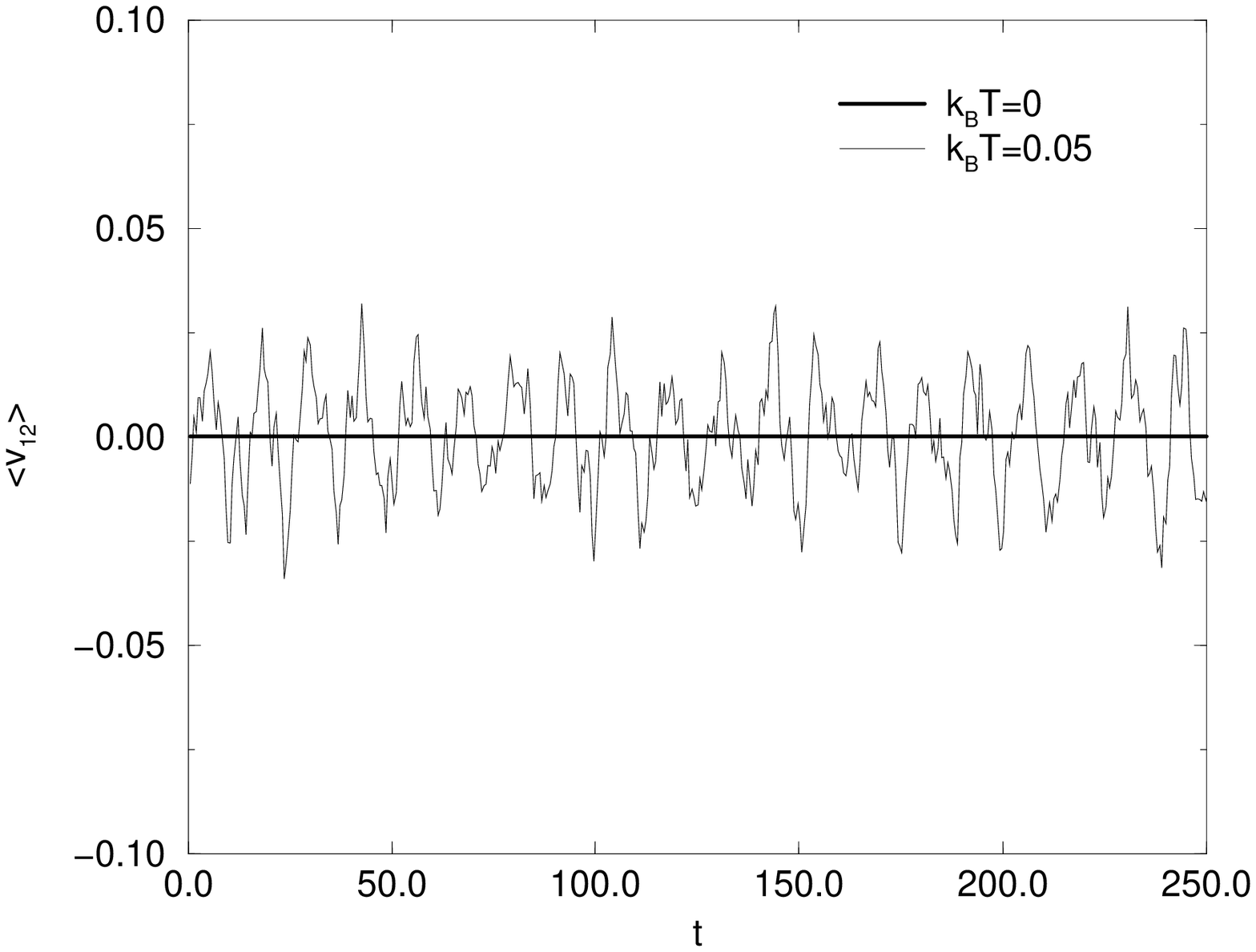}
\leavevmode
\epsfxsize = 3.0in
\epsffile{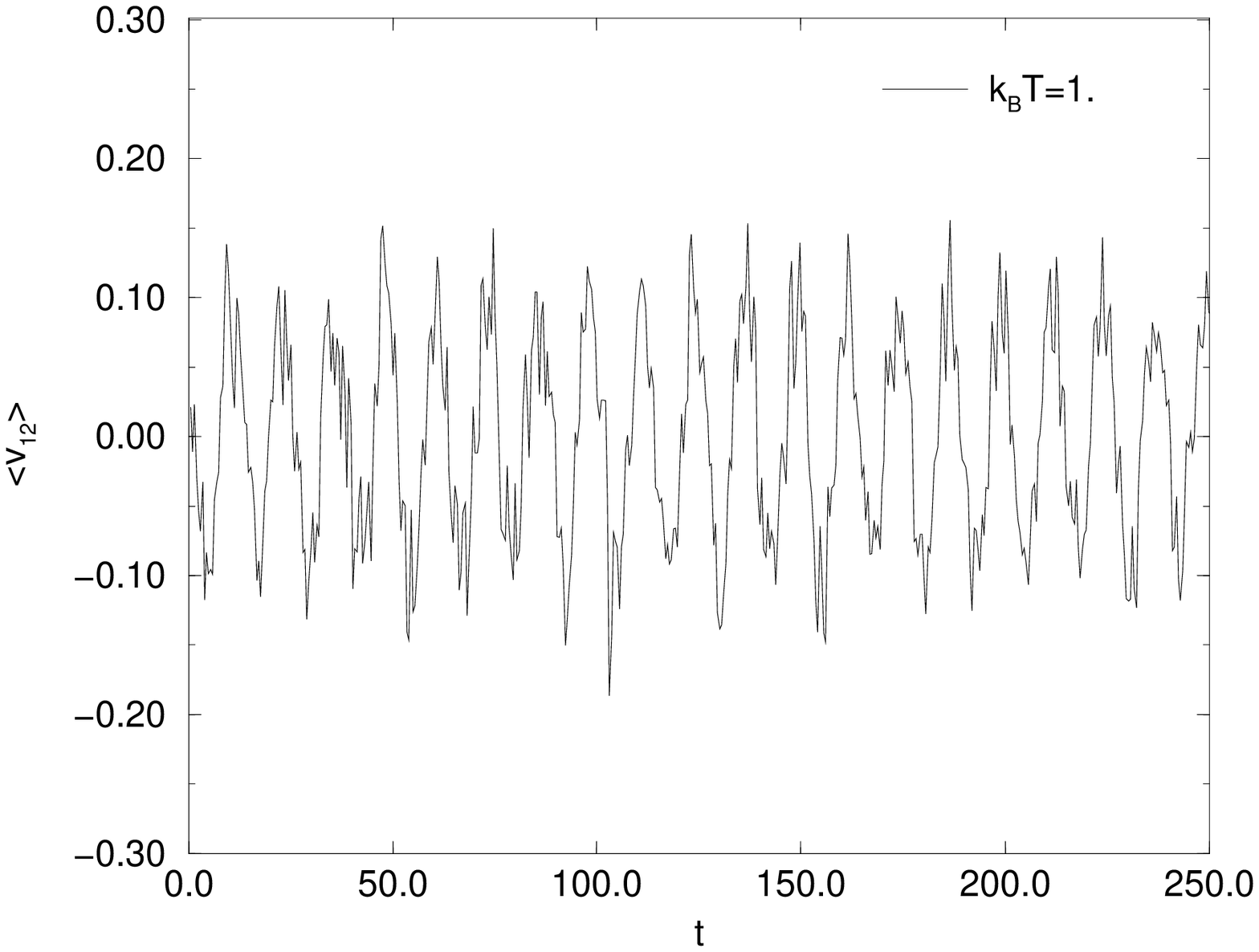}
\vspace{-0.3in}
\end{center}
\begin{center}
\leavevmode
\epsfxsize = 3.0in
\epsffile{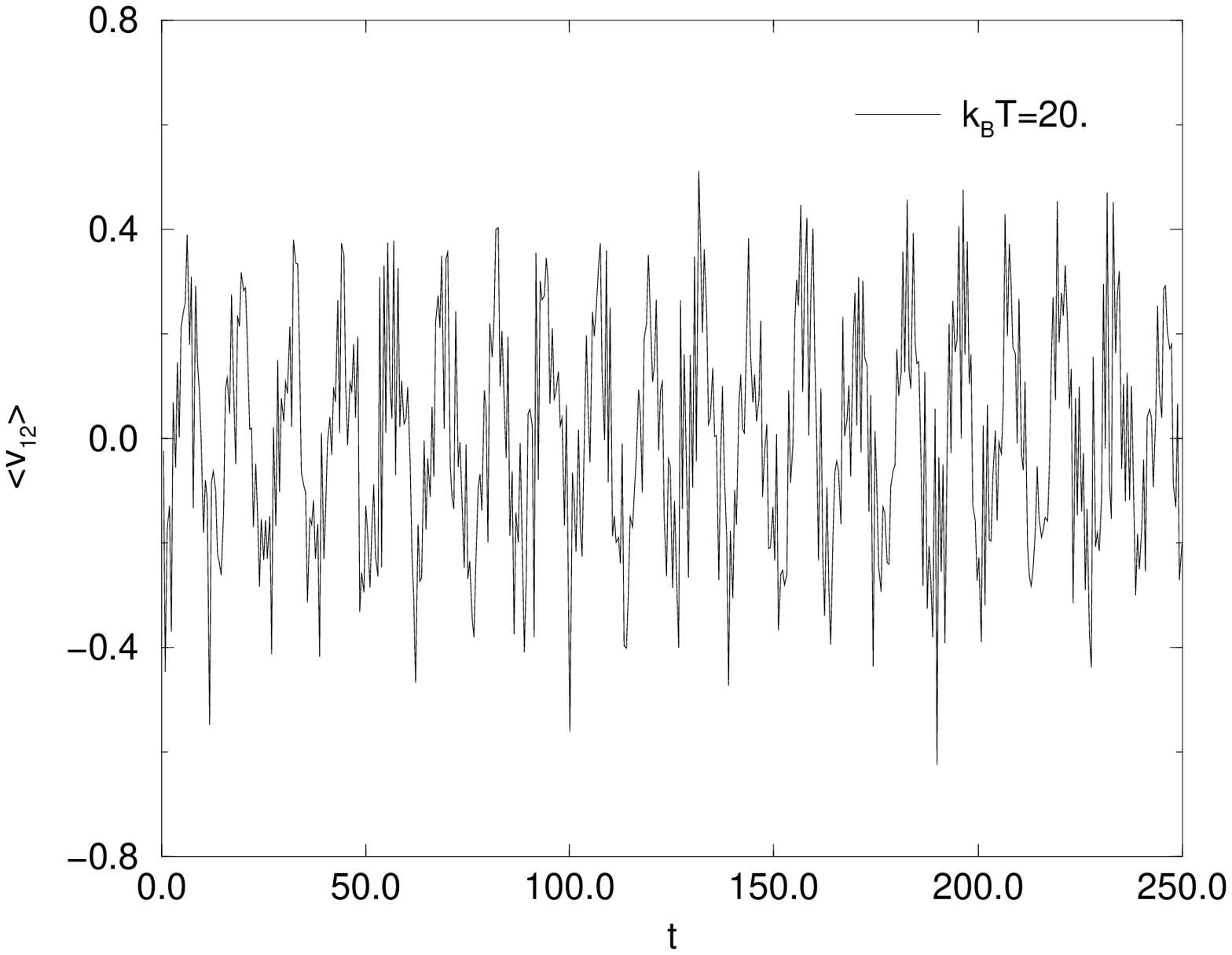}
\leavevmode
\epsfxsize = 3.0in
\epsffile{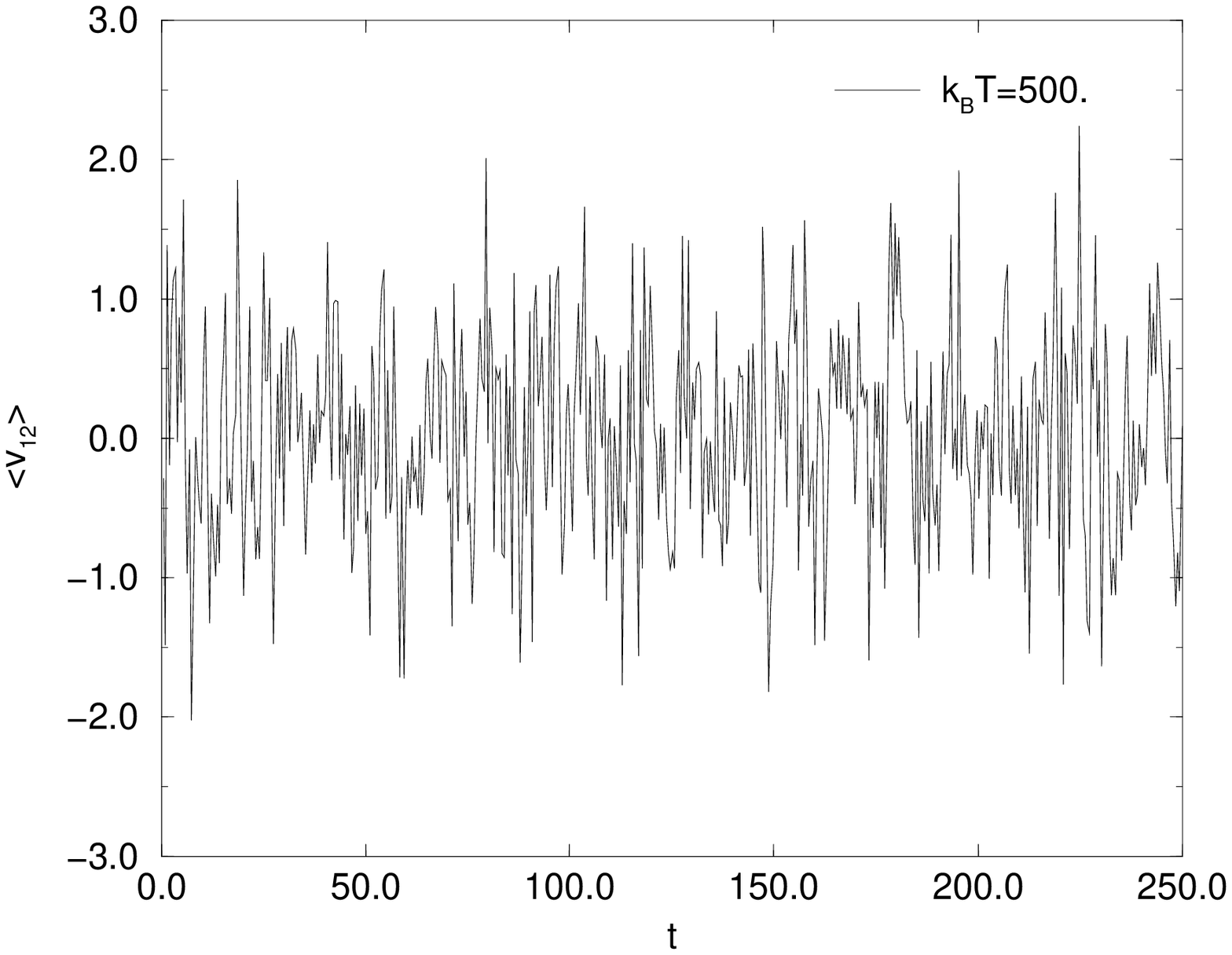}
\end{center}
\caption
{Mean velocity $\langle v\rangle$ of the $12^{th}$ site of a hard anharmonic
chain with $k'=3$, $\gamma=0.5$, $\omega_0=0.5$, and $A=0.5$, for different
values of the temperature. Enhancement of signal propagation is observed with
increasing temperature ($1^{st}$ and $2^{nd}$ panels), but the signal is
destroyed as the temperature further increases ($3^{rd}$ and $4^{th}$ panels).
Note the different vertical scales.}
\label{srhard}
\end{figure}

\begin{figure}[htb]
\begin{center}
\leavevmode
\epsfxsize = 3.0in
\epsffile{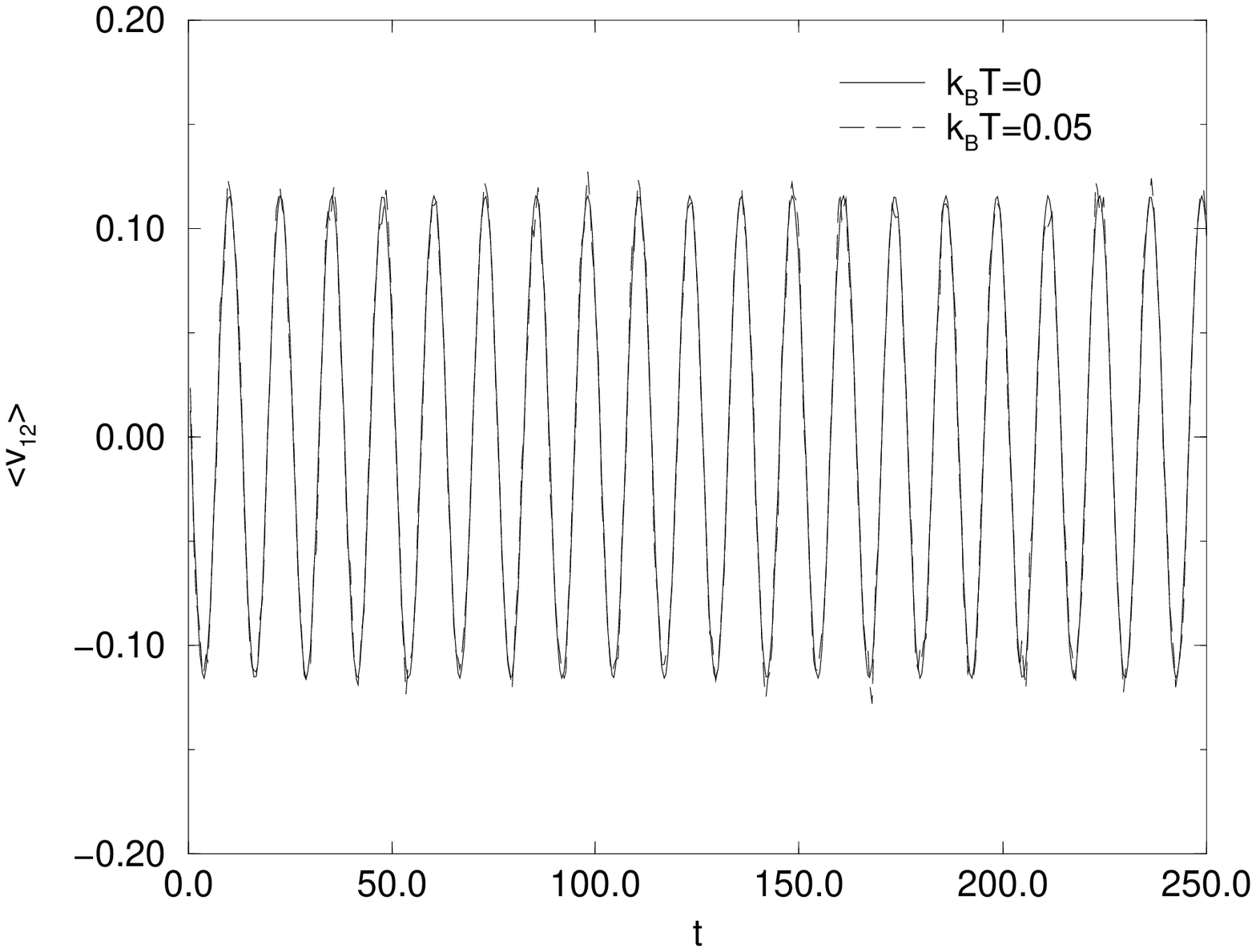}
\leavevmode
\epsfxsize = 3.0in
\epsffile{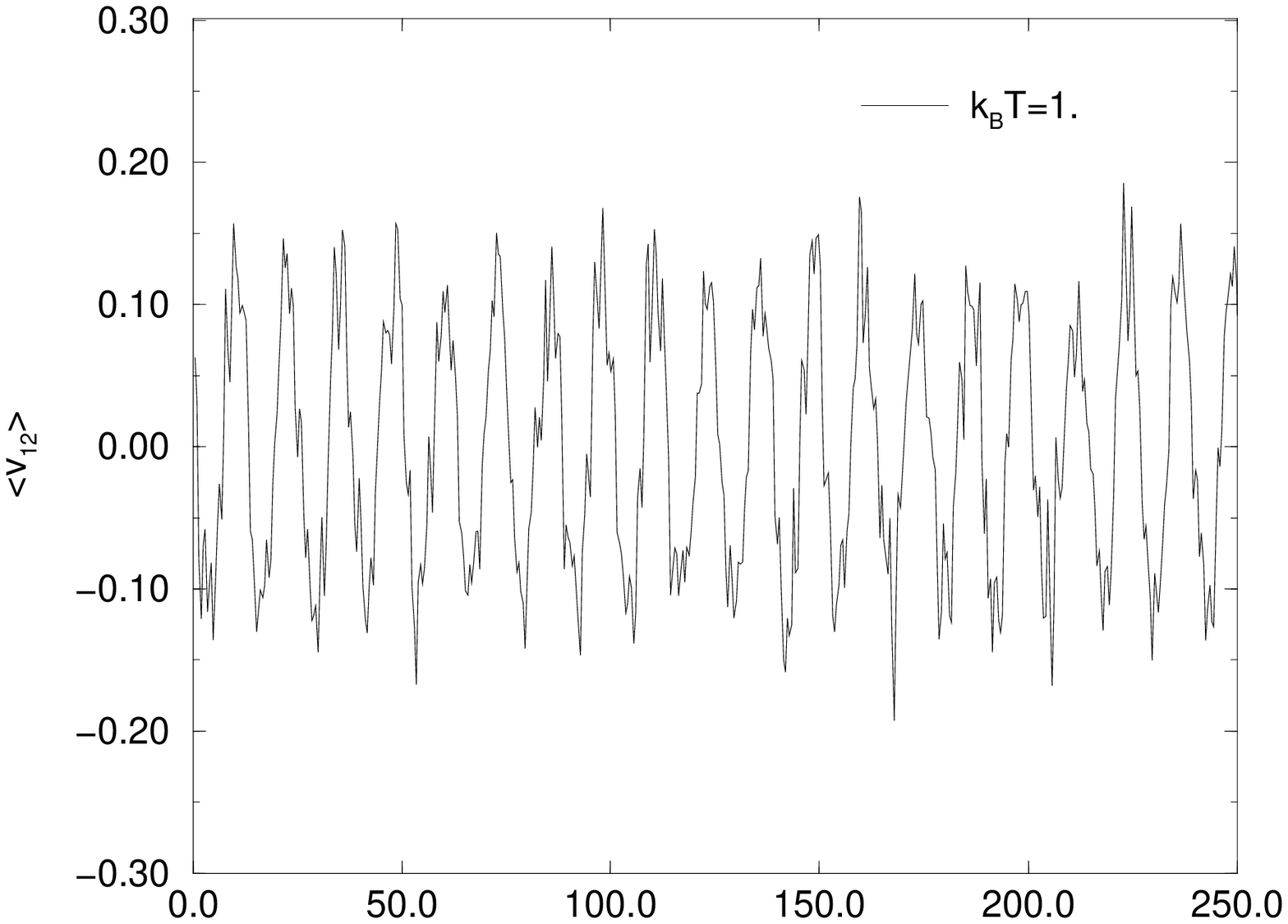}
\vspace{-0.3in}
\end{center}
\begin{center}
\leavevmode
\epsfxsize = 3.0in
\epsffile{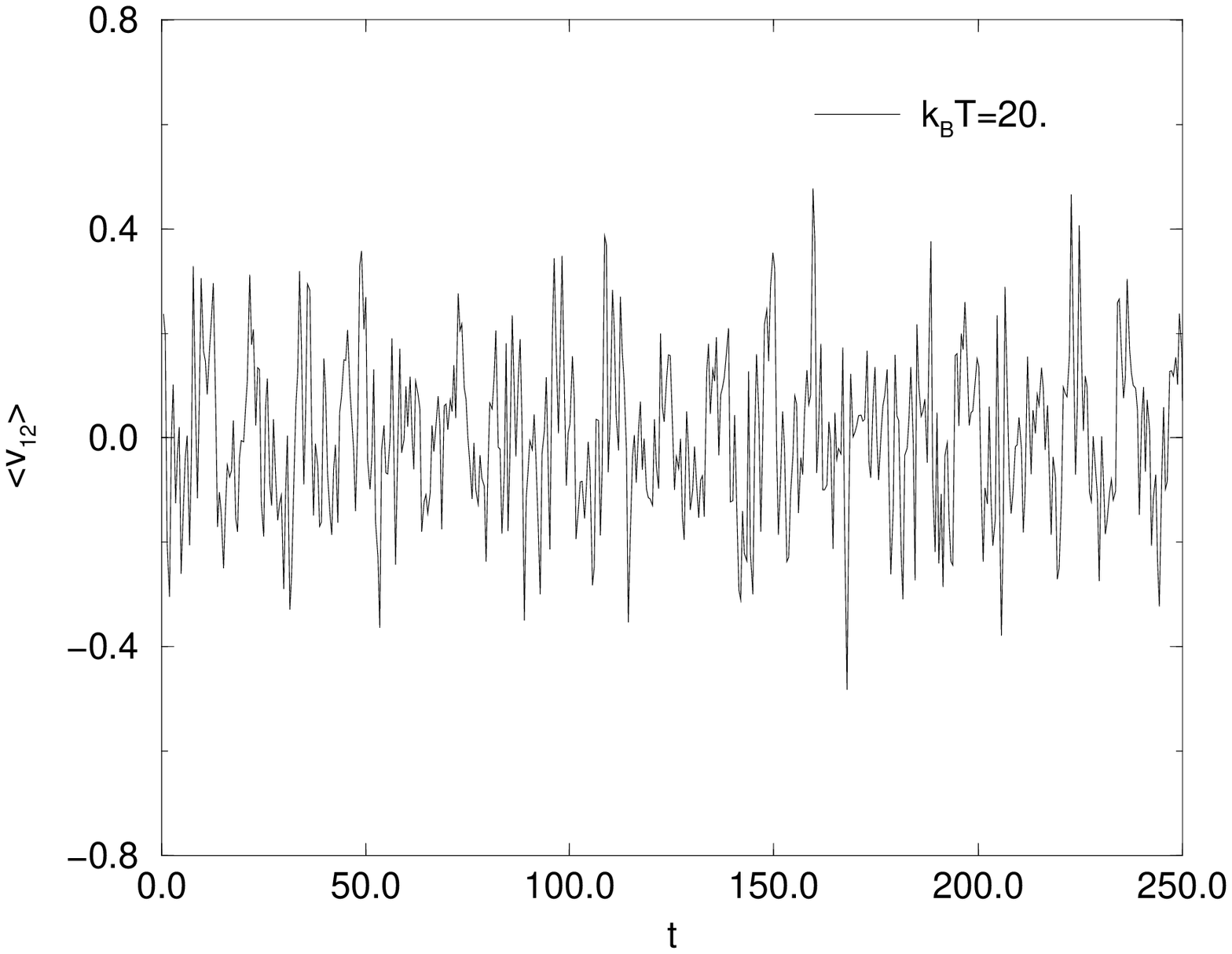}
\leavevmode
\epsfxsize = 3.0in
\epsffile{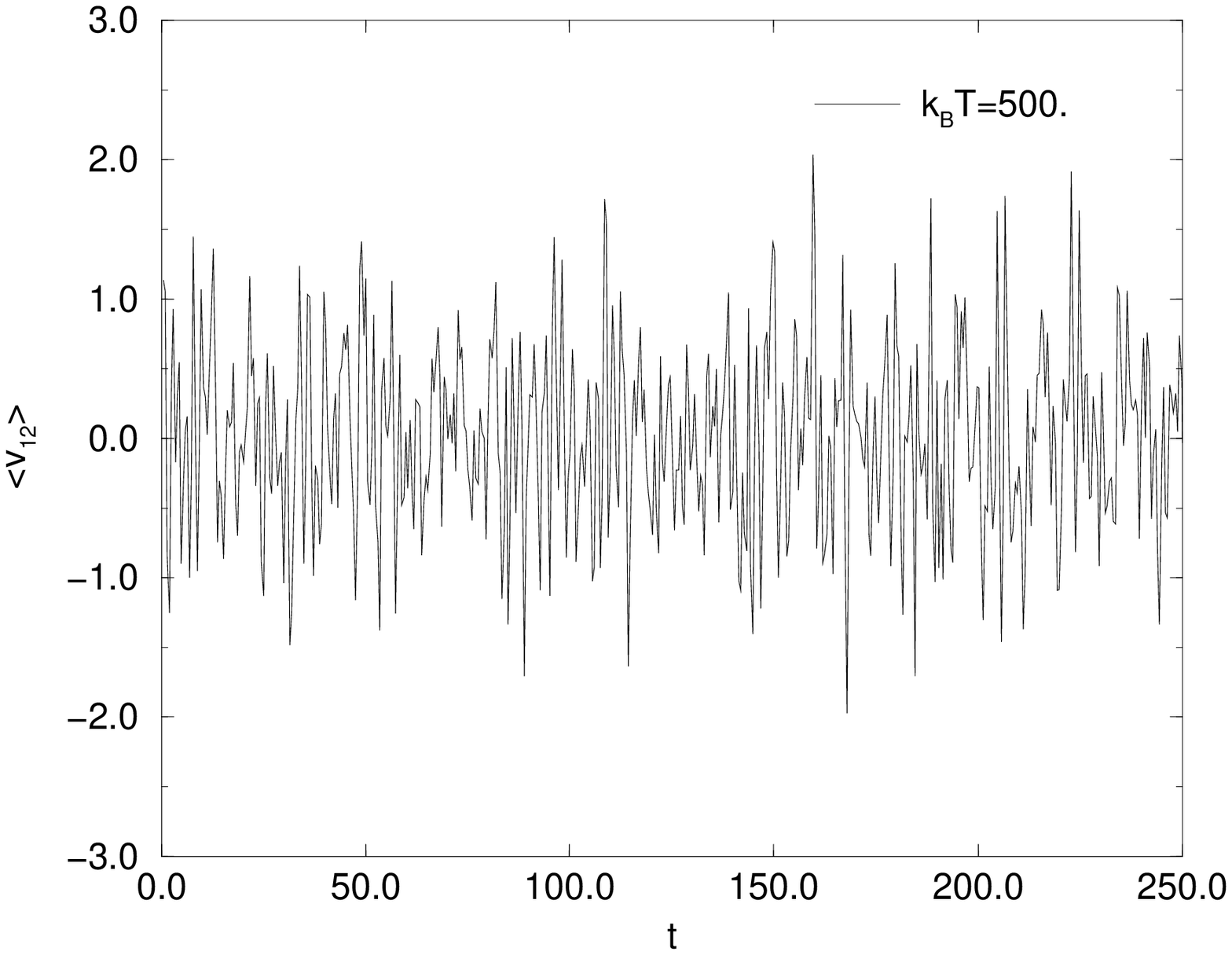}
\end{center}
\caption
{Mean velocity $\langle v\rangle$ of the $12^{th}$ site of the
harmonic
chain with $k=3$, $\gamma=0.5$, $\omega_0=0.5$, and $A=0.5$, for different
values of the temperature. Signal degradation with
increasing temperature is clearly observed.
Note the different vertical scales.}
\label{srharm}
\end{figure}

Fig.~\ref{srhard} is a dramatic but typical demonstration of a realization
of thermal resonance. It shows the mean velocity $\langle v \rangle$
(averaged over realizations) as a function of time for the $12^{th}$ site of a
hard anharmonic chain at different temperatures.
The first panel shows the results for zero
temperature and for a very low temperature;
the temperature increases in
subsequent panels. At zero temperature the $12^{th}$ site hardly moves
because the signal has been dissipated to the bath before reaching this site
(in the corresponding harmonic chain the signal reaches the $12^{th}$ site
quite vigorously at zero temperature, a confirmation of the fact that a given
dissipation is much more effective in a hard anharmonic potential than in a
harmonic one~\cite{our1,our2,Dav} -- see Fig.~\ref{srharm}).
A very small temperature increase (still
first panel) causes a large enhancement of the signal, which clearly now
reaches the site. This is apparent in the oscillatory behavior of the velocity
over and above the noisy background. The temperature in the second panel is
close to its optimal value, that is, the value that most enhances the signal
at this
particular site relative to the thermal background. Hence the motion of
the $12^{th}$ site at this temperature is mostly driven by the periodic
forcing of the first site. A
further increase in the temperature (third panel) causes the average velocity
to become increasingly noisy because ever larger fluctuations dominate the
dynamics. Finally, at a sufficiently high temperature (fourth panel), the
signal is essentially buried in the fluctuations, and the motion is simply
that imposed by the thermal bath. In contrast, in a harmonic array ($k=3$)
the signal at any site simply degrades with increasing
temperature. This is illustrated in Fig.~\ref{srharm}.

\begin{figure}[htb]
\begin{center}
\leavevmode
\epsfxsize = 4.0in
\epsffile{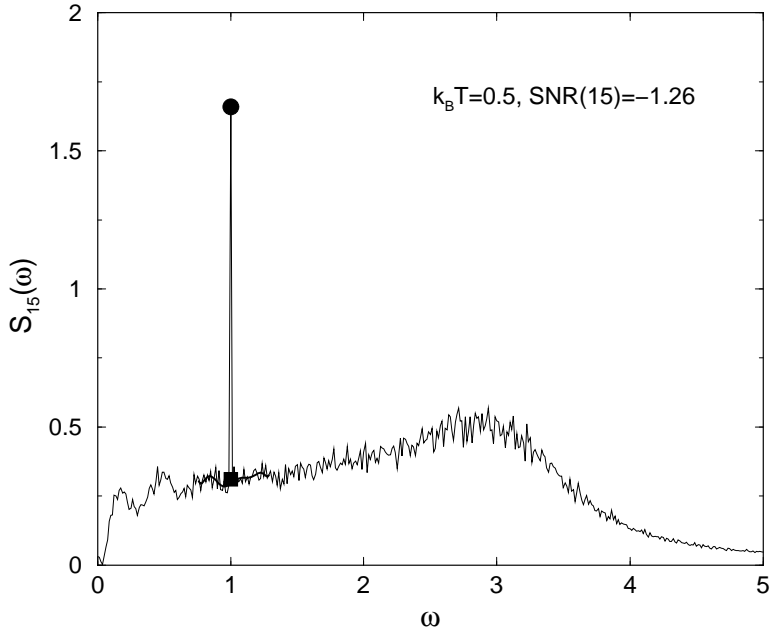}
\end{center}
\caption
{Power spectral density at the $15^{th}$ site of the hard
anharmonic chain with $k'=5$, $\gamma=0.2$, $\omega_0=1.0$ and $A=0.5$ (we
call this our ``standard case" in the text) at temperature $k_BT=0.5$.
The wide solid line shows the polynomial fitting around $\omega_0$.
The circle indicates the value of $S_{15}(\omega_0)$ and the square that
of $S_{15,noise}(\omega_0)$.}
\label{swhard}
\end{figure}

To provide a quantitative measure of the thermal resonance,
we define the power spectral density
$S_j(\omega)$ at each site $j$ as
\begin{equation}
S_j(\omega)  =  \int_{-\infty}^{+\infty} e^{-i \omega \tau} \
\langle v_j(t)v_j(t+ \tau) \rangle \ d \tau~,
\label{sw}
\end{equation}
where the brackets denote an ensemble average over realizations {\em and}
an average over time. In addition to the thermal fluctuations, this function
contains the spectral information about that part of the signal that has
reached site $j$. An example of a portion of the spectrum for the case that
we will call our ``standard case" ($k'=5$, $\gamma=0.2$, $\omega_0=1.0$, $A=0.5$)
is shown in Fig.~\ref{swhard}.
The signal extraction from background noise that
characterizes stochastic resonance is traditionally performed via a
signal-to-noise ratio ($SNR$)~\cite{rev}:
\begin{equation}
SNR(j) \equiv  \log_{10} \Bigg( \frac{{\rm signal~power}~(j)\times \Delta
\omega}
{{\rm thermal~power}~(j)} \Bigg)~,
\label{snr}
\end{equation}
where the signal power is the value
$S_j(\omega_0) - S_{j,noise}(\omega_0)$, the thermal
power $S_{j,noise}(\omega_0)$ is estimated by performing
a $4^{th}$-order polynomial fit to
$S_j(\omega)$ around -- but not including -- the forcing frequency $\omega_0$,
and $\Delta \omega$ denotes the frequency integration step and is equal to
0.0125 (the inverse of the 80 oscillation periods used as our measurement
time) throughout this paper.
This definition of the $SNR$ is not unique, but our results are robust with
respect to variations in this definition.

\section{Thermal Resonances}
\label{results}
\begin{figure}[htb]
\begin{center}
\leavevmode
\epsfxsize = 3.0in
\epsffile{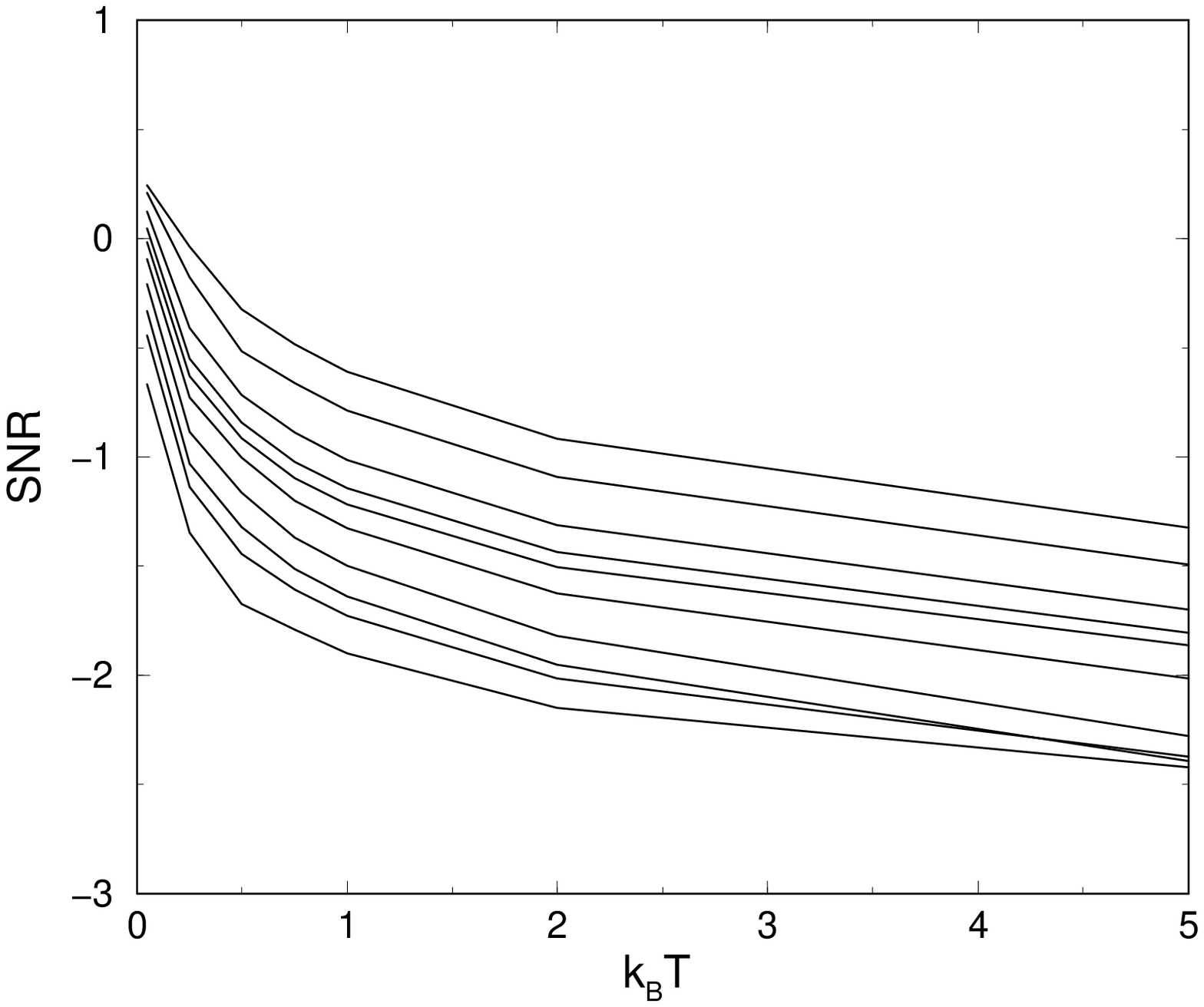}
\leavevmode
\hspace{0.2in}
\epsfxsize = 3.0in
\epsffile{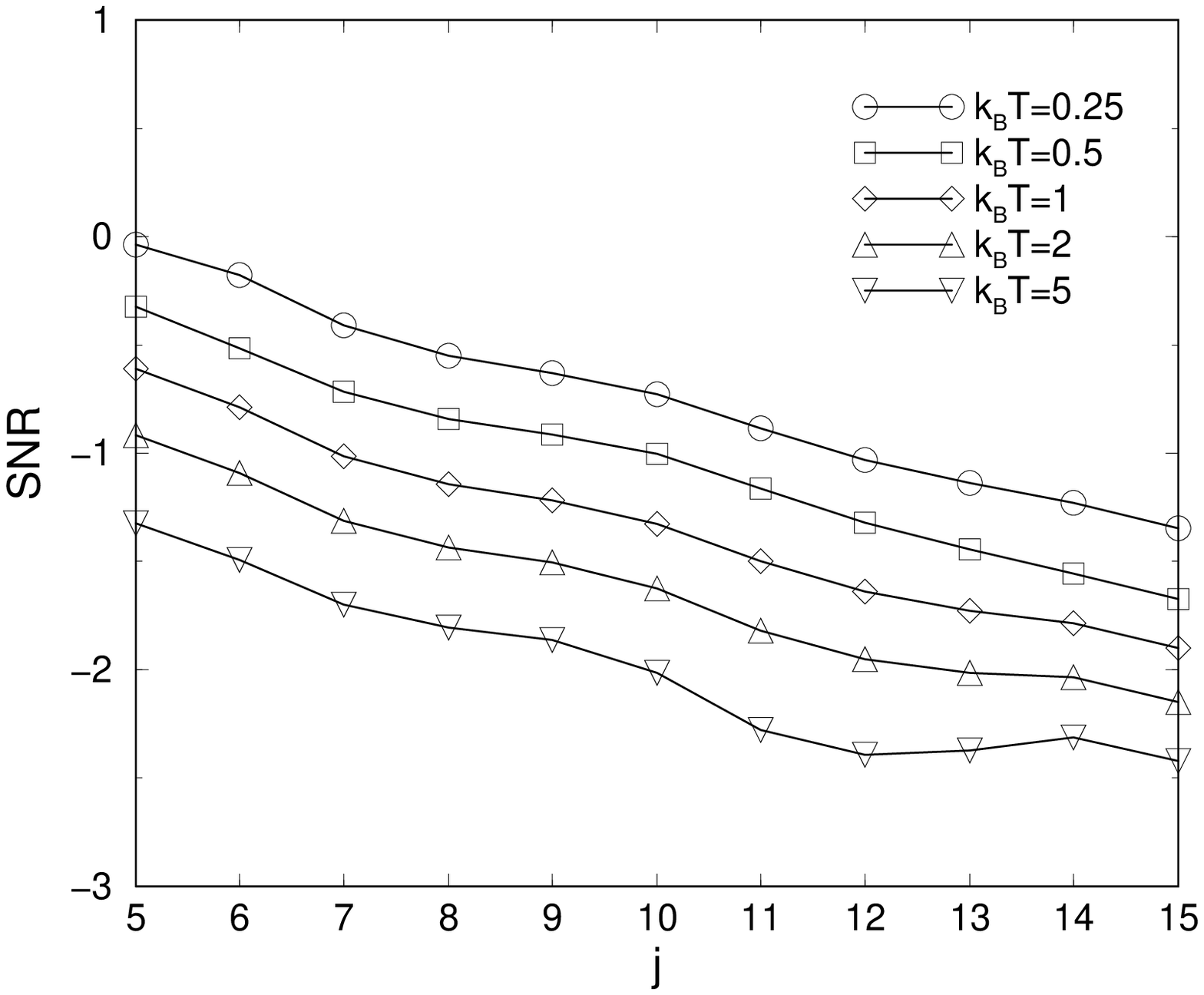}
\end{center}
\caption
{First panel: Typical $SNR$ curves as a function of temperature
for different sites along a harmonic
chain. The $SNR$ for each site decreases
monotonically with temperature.
This particular example shows sites from the $5^{th}$ to the
$14^{th}$ (top to bottom)
with $k=3$, $\gamma=0.5$, $\omega_0=1.0$, and $A=0.5$.
Second panel: $SNR$ as a function of site for different temperatures.}
\label{figsnrosch}
\end{figure}

We first present $SNR$ results for the harmonic chain, so as to
clarify later the ways in which the anharmonic chain behaves differently.
The analytic calculations associated with the harmonic chain are presented
in Appendix~\ref{a}.  These results serve as a test for our
numerical simulations. The first panel in
Fig.~\ref{figsnrosch} shows $SNR$ curves as a function of
temperature for different sites, and the second panel
shows the same results as a function of distance from the forced site
for different temperatures.  The results are exactly as shown in the
Appendix and as one would
expect: the $SNR$ decreases monotonically with increasing
temperature and with increasing distance from the applied
signal.  Note that $SNR(j)$ decreases with increasing temperature
because the numerator in Eq.~(\ref{snr}) is essentially independent
of temperature while the denominator increases (see Appendix~A).
A point to note is that the decay of $SNR(j)$ with $j$ at a given
temperature provides a measure of the shape of the stationary front of the
signal at that temperature.

\begin{figure}[htb]
\begin{center}
\leavevmode
\epsfxsize = 3.0in
\epsffile{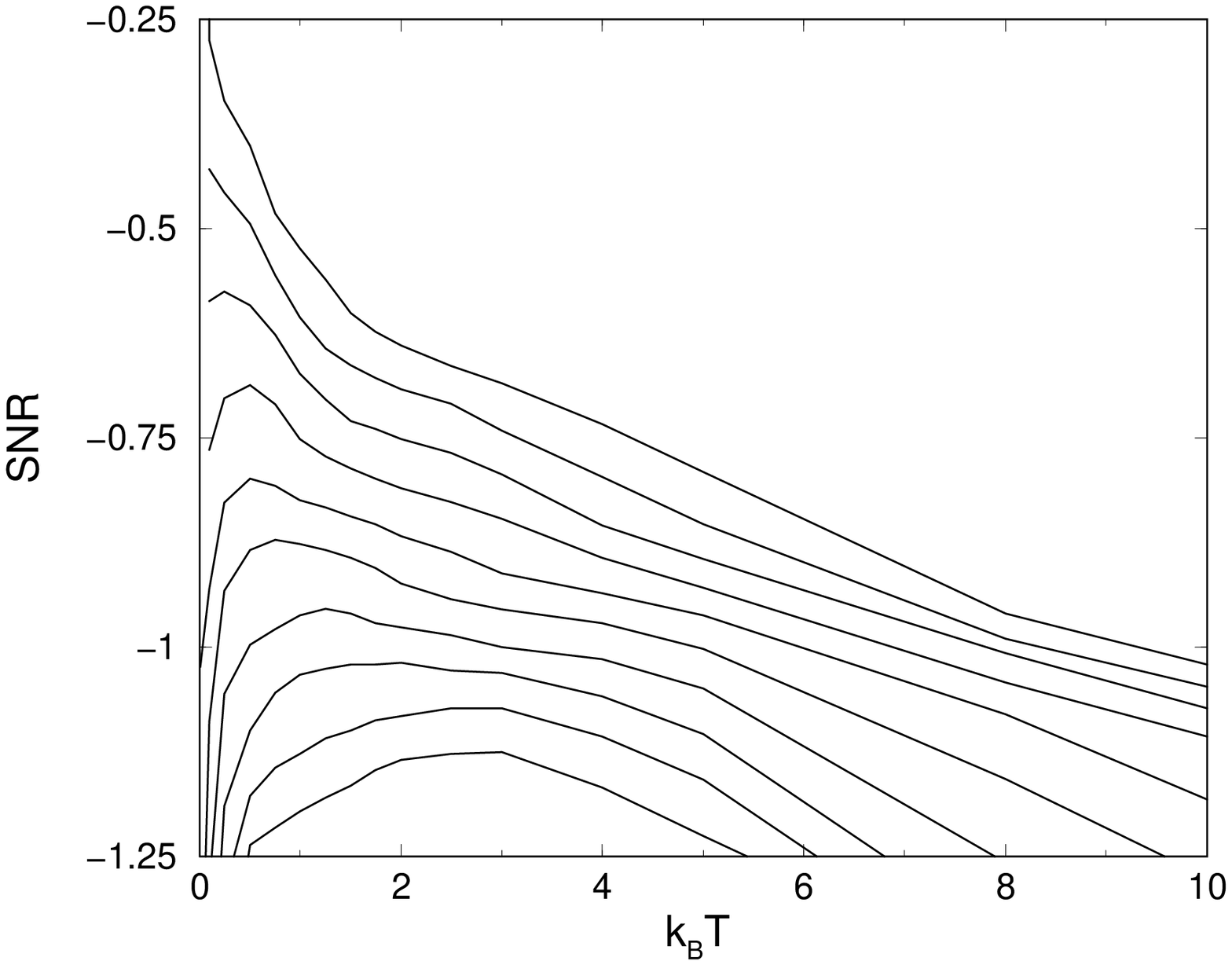}
\leavevmode
\hspace{0.2in}
\epsfxsize = 3.0in
\epsffile{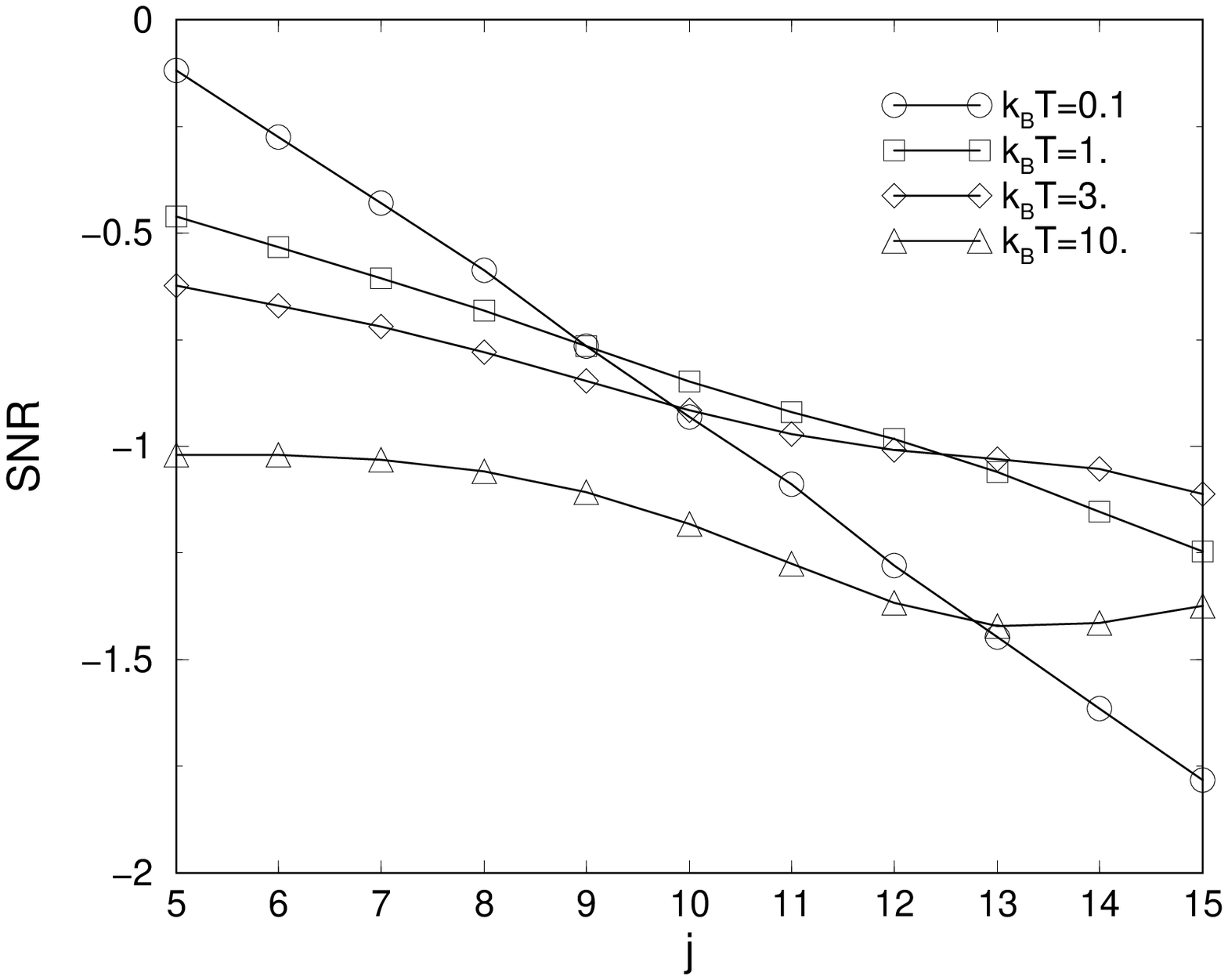}
\end{center}
\caption
{First panel: $SNR$ curves for different sites (from $j=6$ to $15$, top to
bottom)
along the anharmonic
chain as a function of the temperature for the standard case:
$A=0.5$, $k'=5$, $\gamma=0.2$, $\omega_0=1$.
The optimal temperature increases with distance from the first site.
Second panel:
$SNR$ as a function of the site for different temperatures.}
\label{figsnroscd}
\end{figure}

We now turn to the anharmonic chain.
The first panel in Fig.~\ref{figsnroscd} shows $SNR$ curves as
a function of temperature for different sites.
The {\em thermal resonance} is identified with the
$j$-dependent maximum of $SNR(j)$ as a function of temperature. The
signal at the first few sites is monotonically weakened as a function of the
temperature (as it is at {\em all} sites in the harmonic chain), but the $SNR$
is enhanced with increasing temperature for sites farther away until it
reaches a maximum (thermal resonance), beyond which it decreases. Note
that the optimal temperature increases with distance from the first
site. For the particular parameters used in this simulation
the thermal resonance occurs most clearly at temperatures in the
range $k_BT<4$ for sites between the $8^{th}$ and the $15^{th}$; these
details can of course be modified by changing the parameters
(see Section~\ref{parameters}).

Complementary results
for $SNR$ curves as a function of the site for different
temperatures are shown in the second panel in Fig.~\ref{figsnroscd}.
The resonance effects are evidenced by the {\em crossings} of
the different curves.  The
crossings reflect the rise and subsequent drop in the $SNR$ at a given
site, and the fact that the curves cross at different sites confirms
that the thermal resonance temperature varies from site to site.

The non-monotonic behavior of the $SNR$ with temperature
is due to the fact that while the thermal power increases with
$T$, the signal power at first also increases, and more rapidly
than the thermal power.
This is exactly the condition for the
existence of stochastic resonance formulated by Dykman et
al.~\cite{Dykman,Stocks,LRT} for single potentials, but that
theory has not yet been
generalized to extended systems. 
The fast increase of the signal power
is consistent with trends found earlier for transmission of
pulses along discrete arrays~\cite{our2}.  We found that a pulse in a hard
anharmonic chain travels more rapidly with increasing temperature (except
for the lowest temperatures, see below),
while in a harmonic array a change in
temperature has no effect on pulse speed.  While a pulse broadens
with increasing temperature in both arrays, the pulse in the anharmonic
array remains relatively more compact. At sufficiently high
temperatures the signal power at a given site becomes independent
of $T$ (as in the harmonic chain) because the signal response has
reached its maximum value; a further increase in temperature
only affects the signal further down the chain.  The ``crossover"
temperature of the signal power from the increasing to the
saturated behavior depends on the chain and signal parameters,
a dependence explored in the next section.

The fact that the $SNR$ for the sites shown is very low (essentially zero)
for the lowest temperatures at sites beyond the first few shows that for a
purely anharmonic chain the signal essentially stops beyond the first few
sites, whereas in the harmonic example the signal reaches all the sites
shown even at the lowest temperatures shown.  This is also consistent with
the behavior shown earlier for the transmission of a
pulse~\cite{our2}: the pulse velocity at sufficiently low temperatures is
actually lower in the hard chain than in the harmonic
case.  Had we included a harmonic potential contribution equal
to that of the harmonic chain, the $SNR$ vs $k_BT$ curves would
start at the same values as in Fig.~\ref{figsnrosch}, but for
sufficiently distant
sites from the first they would still be non-monotonic.  We have omitted a
harmonic contribution to present the thermal resonance effect in its purest
form.

The apparently monotonic behavior of the first few sites is due to the fact
that for the temperatures shown the signal reaches these sites in any case.
A resonance at these sites would be seen for different parameter values
and/or at even lower temperatures.

\begin{figure}[htb]
\begin{center}
\hspace{0.2in}
\epsfxsize = 3.5in
\epsffile{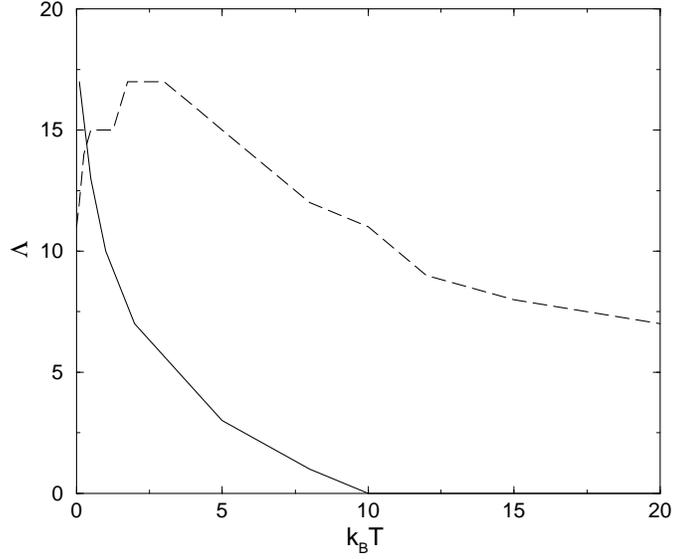}
\end{center}
\caption
{Propagation length $\Lambda$ as a function of the temperature for the
anharmonic chain (standard case) and harmonic chain with $k=1$ (other
parameters as in the standard anharmonic case). The threshold value is
$SNR=-1.2$.}
\label{figjmax}
\end{figure}

Our preceding descriptions point to
another interesting measure of a thermal resonance, namely,
the propagation length
$\Lambda$, defined as the number of sites (i.e., distance along the chain)
for which the $SNR$ exceeds a certain threshold value. Fig.~\ref{figjmax}
presents the temperature dependence of $\Lambda$ for harmonic and anharmonic
arrays and an arbitrarily chosen $SNR$-threshold value of $-1.2$. Since
the harmonic
array does not exhibit thermal resonance, a monotonic decay of $\Lambda$ with
increasing temperature is observed. On the other hand, the hard chain shows a
maximum for a moderate temperature. Again, the
particular values of optimal temperature ($k_BT \approx 3$) and
optimal distance
($\Lambda \approx 17$) can be modified by choosing different parameters
and/or different $SNR$ thresholds, but the qualitative behavior persists
as seen in Fig.~\ref{figjmax}.

\section{Parameter Dependences}
\label{parameters}

The parameters that can be varied in our model are the amplitude
$A$ and frequency $\omega_0$ of the velocity of the first site, the
coupling parameter $k'$, and the damping coefficient $\gamma$.

\begin{figure}[htb]
\begin{center}
\leavevmode
\epsfxsize = 3.0in
\epsffile{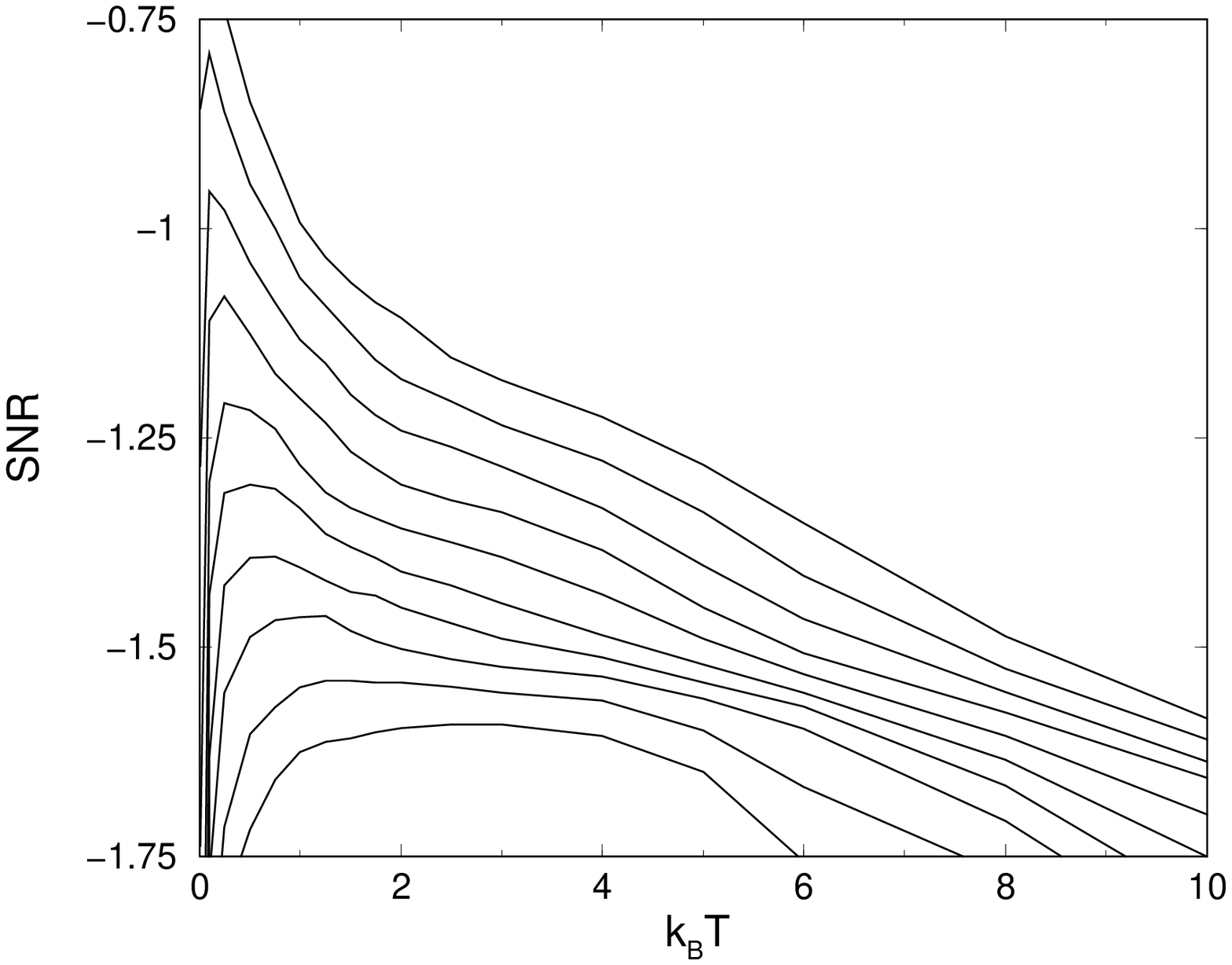}
\hspace{0.2in}
\leavevmode
\epsfxsize = 3.0in
\epsffile{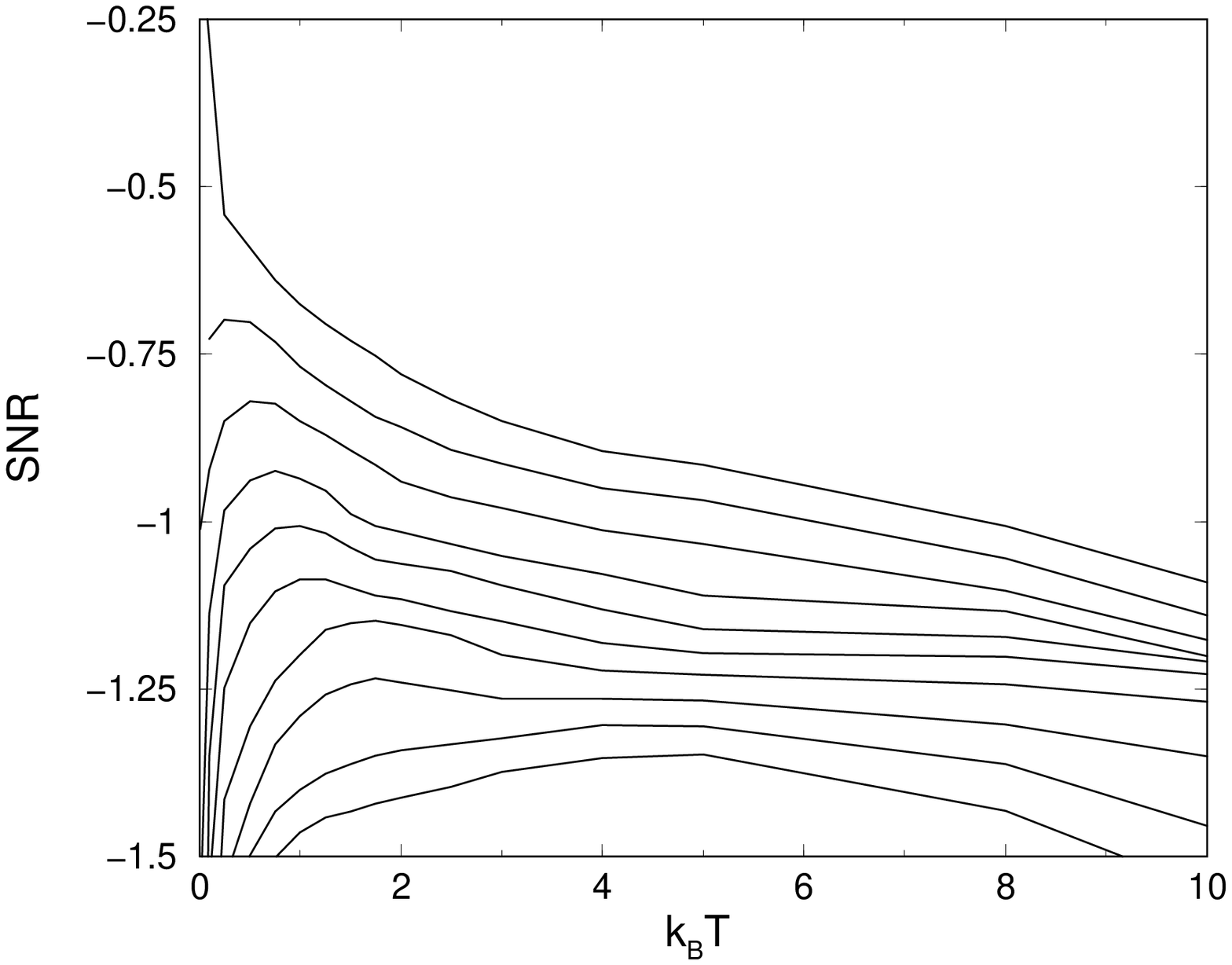}
\end{center}
\caption
{First panel: $SNR$ curves for different sites (from $j=4$ to $13$, top 
to bottom) along the anharmonic
chain as a function of the temperature for the lower-amplitude case:
$A=0.25$, $k'=5$, $\gamma=0.2$, $\omega_0=1$. Second panel:
$SNR$ curves for the weaker-coupling case:
$A=0.5$, $k'=3$, $\gamma=0.2$, $\omega_0=1$.
}
\label{Ak'}
\end{figure}

The first panel of Fig.~\ref{Ak'} shows the $SNR$ curves as a function
of temperature for
a lower amplitude than in the standard case.  Comparison with the first
panel of Fig.~\ref{figsnroscd} shows that the overall $SNR$ is now (of
course) lower
and that the resonance temperature at each site has increased.
The latter behavior indicates that for lower amplitudes the
crossover temperature from an increasing signal to a saturated
signal increases as the signal weakens.
The second panel shows results for weaker coupling.
Again the overall $SNR$ is lower and the resonance temperature higher
at each site.  A weaker coupling thus has the effect of weakening
the signal.

\begin{figure}[htb]
\begin{center}
\leavevmode
\epsfxsize = 3.0in
\epsffile{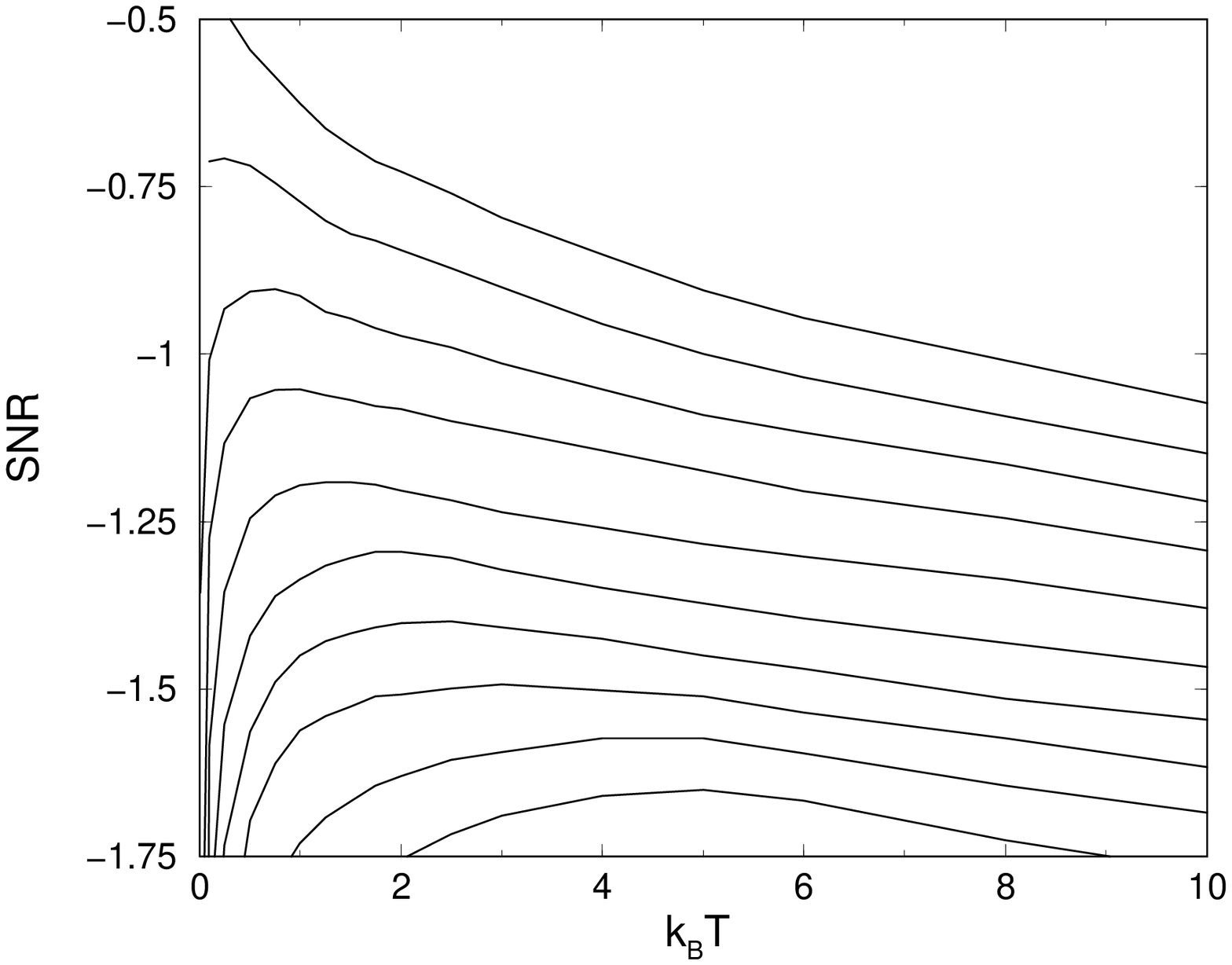}
\leavevmode
\hspace{0.2in}
\epsfxsize = 3.0in
\epsffile{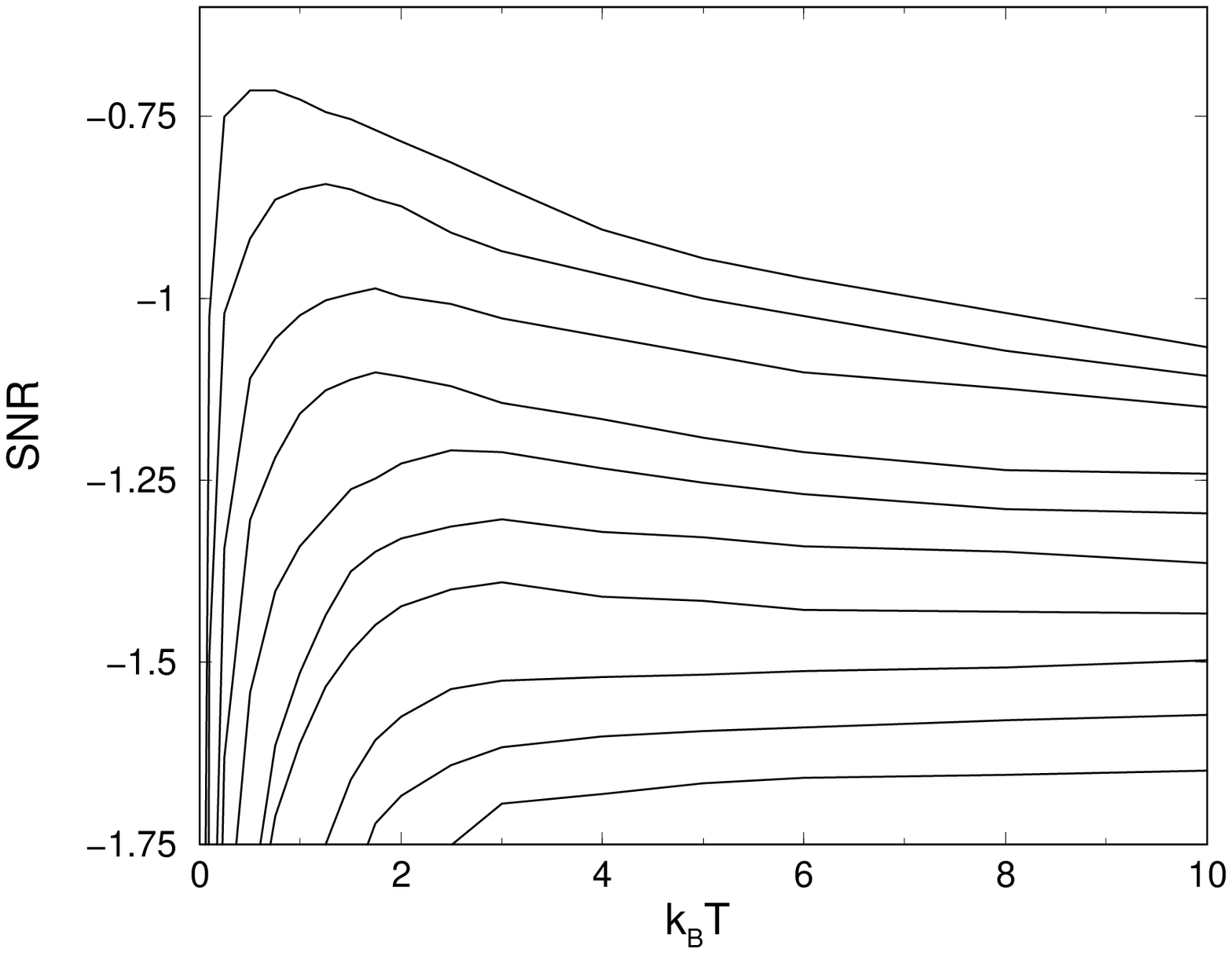}
\end{center}
\caption
{$SNR$ curves for sites 4 to 13 (top to bottom)
for the higher damping case:
$A=0.5$, $k'=5$, $\gamma=0.5$, $\omega_0=1$. Second panel:
$SNR$ curves for the higher frequency case:
$A=0.5$, $k'=5$, $\gamma=0.2$, $\omega_0=2$.
}
\label{gammaomega}
\end{figure}

The first panel of Fig.~\ref{gammaomega}
shows the $SNR$ curves as a function of temperature for higher
damping.  Comparison with the first
panel of Fig.~\ref{figsnroscd} shows that the overall $SNR$ is again
lower than in the standard case and the resonance temperature higher.
The same trends are observed with a higher driving frequency as shown in
the second panel of Fig.~\ref{gammaomega}.  Again, each of these
changes leads to an effectively weakened signal.

Figures~\ref{Ak'} and \ref{gammaomega} illustrate clear trends.
Decreasing $A$ or $k'$, or increasing $\gamma$ or
$\omega_0$, lead to the following consequences: a) the $SNR$
at any given site and temperature decreases; b) the
$SNR$ resonance temperature increases at any given site; c) at a given
temperature the resonance occurs at a site closer to the first.
These trends are consistent with those found earlier for transmission of
pulses along similar arrays~\cite{our2}.  We found that a pulse in a hard
anharmonic chain travels more rapidly and relatively more compactly
with increasing amplitude, decreasing damping, increasing coupling, or
increasing driving frequency.  By contrast, in a harmonic array changes
in temperature or in the signal and damping parameters have no effect on pulse
speed.

\section{Conclusion}
\label{conclusion}

We have demonstrated new
thermal resonances in simple one-dimensional arrays of masses
connected by hard springs. Our chains are not overdamped and hence include
inertial effects. We have shown that the distance and amplitude of
propagation of a signal imposed at one end of the chain can each
be optimized by tuning the temperature of the system.
The resonance behavior reflects the
temperature dependence of the distance traveled by the front of
a signal and the dispersion of this front once steady state has
been reached.  At a fixed temperature,
any parameter change that leads to an increase in velocity of propagation
tends to increase the range of sites where the signal can be detected
as well as
the $SNR$ at a given site. Increased velocity of propagation is also
associated with a lowering of the resonance temperature for a given site.

A complete understanding of the behavior of these simple
anharmonic arrays in a thermal environment requires and relies on
a number of other
inquiries, some of which we have undertaken.  One concerns the
distribution of energy, and the persistence and mobility of energy
fluctuations, in thermal equilibrium~\cite{our1,our3}.  We have found
that soft anharmonic chains (that is, chains with soft interaction
potentials) experience greater energy fluctuations than harmonic
chains, which in turn experience greater fluctuations than hard
chains.  This is a direct consequence of the virial theorem.
We have also established that fluctuations are mobile in harmonic
and hard
chains but not in the soft.  Most interestingly, thermal
fluctuations travel most rapidly and remain localized over
considerably greater distances in the hard chain.  These results
in turn can lead to very different transition rate statistics and
effective damping coefficients for a bistable impurity embedded in
each of these arrays at a given temperature~\cite{our3}.

Another interesting line of inquiry concerns the propagation of
an energy pulse along such arrays. As mentioned earlier, we have
found that
the propagation of an energy pulse in a hard anharmonic array
can be enhanced by immersing the array in a thermal bath, and
that hard anharmonicity in the springs causes a tight and
persistent packing of the energy~\cite{our2}.

Almost all of our results are numerical.  Analytic results for the
harmonic array provide some insights and point toward the possibility
that an approach based on linearization methods~\cite{linearization}
together with a linear
response theory~\cite{LRT} might lead to analytic insights for
nonlinear arrays.  We are currently exploring these possibilities.

The thermal resonances and other interesting behavior that we have
found for simple hard anharmonic chains are likely to be
prototypical and therefore applicable to many other discrete
systems with anharmonic interactions.  While many other systems
have been investigated in which a careful balance of parameters at
zero temperature or a manipulation of external noise lead to
interesting localization, resonance, and synchronization phenomena,
our systems are among the simplest
generic systems in which the temperature can be
used as the tuning parameter to achieve similar effects.

\section*{Acknowledgments}

R. R. gratefully acknowledges the support of this research by the
Ministerio de Educaci\'{o}n y Cultura through Postdoctoral Grant No.
PF-98-46573147.  A. S. acknowledges sabbatical support from DGAPA-UNAM.
This work was supported in part by the Engineering Research Program of the
Office of Basic Energy Sciences at the U.S. Department of Energy under
Grant No. DE-FG03-86ER13606.

\appendix
\section{Harmonic Chain}
\label{a}
In this appendix we consider a harmonic chain and calculate analytically
some of the numerical results presented or invoked in the body of the paper.
Although the behavior of linear chains is in general well understood, it is
useful to present results in the particular context of this
work.  These are not readily available in the literature.

\subsection{Zero Temperature, With Signal}
\label{appa}

Consider first an infinite linear chain ($-\infty < n < \infty$)
at zero temperature, that is, in the
absence of fluctuations (but with the dissipative contribution).  The
equations of motion then are
\begin{equation}
\ddot x_n   =   k(x_{n+1} +x_{n-1} - 2x_n) - \gamma \dot{x}_n,
\label{alang}
\end{equation}
with the signal $\dot{x}_0  = A  \sin ( \omega_0 t)$ applied at site
$n=0$.
We conjecture the quasi-stationary solution
\begin{equation}
x_n^s(t) = -\frac{A}{\omega_0}e^{-\mu |n|} \cos (\omega_0t +|n|b) +c_n
\label{solution}
\end{equation}
where the superscript $s$ denotes the presence of the signal.
The constants $\mu$ and $b$ are to be determined.  The additive constant
$c_n$ must be independent of $n$ -- it simply represents an overall
translation of the chain (because it is not anchored) but otherwise does
not contribute to the velocity analysis.  The velocities associated with
Eq.~(\ref{solution}) are
\begin{equation}
\dot{x}_n^s(t) = Ae^{-|n|\mu} \sin (\omega_0t +|n|b).
\label{solutionv}
\end{equation}
This result is clearly consistent with the imposed signal at $n=0$.
The solution (\ref{solutionv}) can be used to calculate the
zero-temperature signal according to Eq.~(\ref{sw}):
\begin{equation}
S_n^s(\omega_0)  =  \int_{-\infty}^{+\infty} e^{-i \omega_0 \tau} \
\langle \dot{x}_n(t)\dot{x}_n(t+ \tau) \rangle \ d \tau~.
\label{asw}
\end{equation}
The average (which in this deterministic case is only over time)
eliminates rapidly oscillating contributions,
leaving only a $\delta$-function-type of contribution.
The result is the power spectral density
\begin{equation}
S_n^s(\omega_0) = \frac{A^2}{2} e^{-2|n|\mu}.
\label{asignal}
\end{equation}

To find the constant $\mu$ we substitute
(\ref{solution}) into (\ref{alang}) and set the
coefficients of $e^{i\omega_0t}$ and those of $e^{-i\omega_0t}$ equal
to zero (one resulting equation is simply the complex
conjugate of the other).  This immediately leads to the relation
\begin{equation}
-\omega_0^2 = k (e^{-\mu +ib} + e^{\mu-ib} -2) -i\gamma\omega_0.
\end{equation}
We make the substitution $u\equiv e^{-\mu+ib}$
and note the symmetry in $u$ and $1/u$:
\begin{equation}
k\left(u+\frac{1}{u} -2\right) +\omega_0^2 -i\gamma\omega_0 =0.
\end{equation}
Next we multiply through by $u$ and solve the resulting quadratic equation:
\begin{equation}
u_{\pm} = \frac{-(\omega_0^2-2k -i\gamma\omega_0)\pm \sqrt{(\omega_0^2 -2k
-i\gamma\omega_0)^2 -4k^2}}{2k}.
\label{careful}
\end{equation}
Note that $u_+u_-=1$, so if one solution
is $e^{-\mu+ib}$, then the other is $e^{\mu-ib}$.  Which is which is not
clear at this point. $\mu$ must be positive for a physically
acceptable solution, and this will be used below to sort out the
choice.

To extract $\mu$
we calculate the ratio $u_+/u_-^*$, which is either $e^{2\mu}$ or
$e^{-2\mu}$.  If $u_+=e^{-\mu+ib}$ then $u_+/u_- = e^{-2\mu}$.
If, on the other hand, $u_+=e^{\mu-ib}$ then $u_+/u_- = e^{2\mu}$.
Which it is will be seen at the end when we inspect the magnitude of the
result.

In calculating the ratio $u_+/u_-^*$ one must exercise caution
in taking complex conjugates because of the square root in
Eq.~(\ref{careful}); it is not appropriate to simply change every $i$ to
a $-i$ in such an expression if one does not know the signs of the
terms inside the square root.
It is helpful
to write each complex number in terms of an amplitude and a phase:
\begin{equation}
(\omega_0^2-2k -i\gamma\omega_0) \equiv 2k\alpha e^{i\beta}
\end{equation}
so that
\begin{eqnarray}
2k\alpha &=& \left[ (\omega_0^2 -2k)^2 +\gamma^2\omega_0^2\right]^{1/2},
\nonumber\\ [12pt]
\beta &=& \tan^{-1} \left(\frac{-\gamma\omega_0}{\omega_0^2-2k}\right).
\end{eqnarray}
Similarly, with
\begin{equation}
(\omega_0^2 -2k -i\gamma\omega_0)^2 -4k^2 \equiv 4k^2 \nu^2 e^{2i\epsilon}
\end{equation}
we find
\begin{eqnarray}
4k^2\nu^2 &=& \omega_0 \left[ \omega_0^2(\omega_0^2-4k-\gamma^2)^2 +4\gamma^2
(\omega_0^2 -2k)^2\right]^{1/2},
\nonumber\\ [12pt]
2\epsilon &=& \tan^{-1} \frac{-2\gamma(\omega_0^2-2k)}
{\omega_0(\omega_0^2 -4k -\gamma^2)}.
\end{eqnarray}
Then
\begin{equation}
u_+ = -\alpha e^{i\beta} +\nu e^{i\epsilon}, \qquad
u_- = -\alpha e^{i\beta} -\nu e^{i\epsilon},\qquad
\end{equation}
and the ratio $u_+/u_-^*$ is thus real.  Multiplying
top and bottom of this ratio by $u_-$ gives
\begin{equation}
\frac{u_+}{u_-^*} = \frac{u_+u_-}{u_-^*u_-} =\frac{1}{|u_-|^2}.
\end{equation}
Further and explicitly
\begin{equation}
|u_-|^2 = (-\alpha e^{i\beta} -\nu e^{i\epsilon})\times
(-\alpha e^{-i\beta} -\nu
e^{-i\epsilon}) = \alpha^2 +\nu^2 +2\alpha\nu \cos (\beta-\epsilon).
\end{equation}

All that remains is the evaluation of the resulting expressions.  In
summary
\begin{equation}
\alpha^2 +\nu^2 +2\alpha\nu \cos (\beta-\epsilon) = e^{-2\mu}
\end{equation}
or
\begin{equation}
\alpha^2 +\nu^2 +2\alpha\nu \cos (\beta-\epsilon) = e^{+2\mu}.
\end{equation}
The choice is determined by whether the result is $<1$ (in which case
the first equality holds) or $>1$ (in which case it is the second).
The quantities $\alpha$, $\nu$, $\beta$, and $\epsilon$ are
thus completely defined in terms of the system parameters.

We have tested our numerical simulations against the prediction
(\ref{asignal}) for large ranges of parameter values and have found
agreement to five significant figures.

The solutions presented in this section do not obey $N$-periodic boundary
conditions.  It is fairly straightforward using an image-like method to
construct explicitly periodic solution by applying the signal at
$0,\pm N,\pm 2N,\dots$ but the changes would be exponentially small in $N$
and unimportant for sufficiently long chains.  The quasi-stationary
behavior assumed above for sites not too distant from $n=0$ sets in long
before the signal at $n=0$ reaches sites $\pm N$.

\subsection{Finite Temperature, No Signal}

Now we solve the equations of motion
\begin{equation}
\ddot x_n   =   k(x_{n+1} +x_{n-1} - 2x_n) - \gamma \dot{x}_n + f_n(t)
\label{blang}
\end{equation}
without a signal.
We are again interested in the stationary behavior.

We define the Fourier transform and its inverse,
\begin{equation}
y_q = \sum_{n=0}^{N-1} x_n e^{2\pi iqn}, \qquad \qquad
x_n = \frac{1}{N}\sum_{q=0}^{N-1} y_q e^{-2\pi i nq}.
\end{equation}
Transforming Eq.~(\ref{blang}) immediately leads to
\begin{equation}
\ddot{y}_q +
4 k \sin^2\left( \frac{\pi q}{N} \right) y_q + \gamma
\dot{y}_q
\ = \ F_q(t)
\label{ftheequation}
\end{equation}
where the inhomogeneous term is the transform of the noise:
\begin{equation}
F_q(t)\equiv \sum_{n=0}^{N-1} f_n(t)e^{2\pi iqn}.
\end{equation}
The solution of this second order inhomogeneous differential equation
is of standard form.  The initial conditions (which we will take to be
zero) are unimportant since we seek the long-time behavior:
\begin{equation}
y_q^0(t)  \ = \ \lim_{t\rightarrow\infty} \
\frac{1}{[r_2(q) -r_1(q)]} \int_0^t F_q(\tau)
\left( e^{r_2(q) (t-\tau)} -e^{r_1(q)(t-\tau)}\right) d\tau
\label{thisone}
\end{equation}
and consequently
\begin{equation}
\dot{y}_q^0(t) \ = \ \lim_{t\rightarrow\infty} \
\frac{1}{[r_2(q) -r_1(q)]} \int_0^t
F_q(\tau) \left( r_2(q)e^{r_2(q) (t-\tau)}
-r_1(q)e^{r_1(q)(t-\tau)}\right) d\tau,
\end{equation}
where
\begin{equation}
r_{1,2}(q) = -\frac{\gamma}{2} \pm \sqrt{\left(\frac{\gamma}{2}\right)^2
- 4k\sin^2\left(\frac{\pi q}{N}\right)}.
\end{equation}
The superscript $0$ is used to stress the absence of a signal.

With this result and the correlation function for the thermal fluctuations
that follows immediately from Eq.~(\ref{fdr}),
\begin{equation}
\left< F_q(t_1)F_{q'}(t_2)\right> \ = \
2 \gamma \ k_BT \ \delta(t_1 -t_2) N \delta_{q,-q'},
\end{equation}
we obtain
upon integration and Fourier inversion the velocity
correlation function
\begin{eqnarray}
C_n^0(\tau) \ &\equiv& \ \left< \dot{x}_n(t) \dot{x}_n(t+\tau)\right>
\nonumber \\ [12pt]
\ & =& \ \frac{\gamma k_BT}{N} \ \sum_{q=0}^{N-1} \
\frac{1}{[r_2^2(q)-r_1^2(q)]} \left( r_1(q)e^{r_1(q)t} -
r_2(q)e^{r_2(q)t}\right).
\end{eqnarray}
We note in passing that this reduces to the standard correlation function
for a single harmonic oscillator in a heat bath, $C_n^0(\tau)\rightarrow
k_BT e^{-\gamma\tau}$,
when the coupling
coefficient $k\rightarrow 0$.  The associated power spectral density is
\begin{eqnarray}
S_n^0(\omega) \ &=& \
2\int_0^\infty d\tau \ C_n^0(\tau)
\cos (\omega\tau)
\nonumber\\ [12pt]
\ &=& \
\frac{2\gamma k_BT\omega^2}{N} \ \ \sum_{q=0}^{N-1} \
\frac{1} {[r_1 ^2(q)+\omega^2] [r_2 ^2(q)+\omega^2]}.
\label{predb}
\end{eqnarray}

\begin{figure}[htb]
\begin{center}
\leavevmode
\epsfxsize = 3.5in
\epsffile{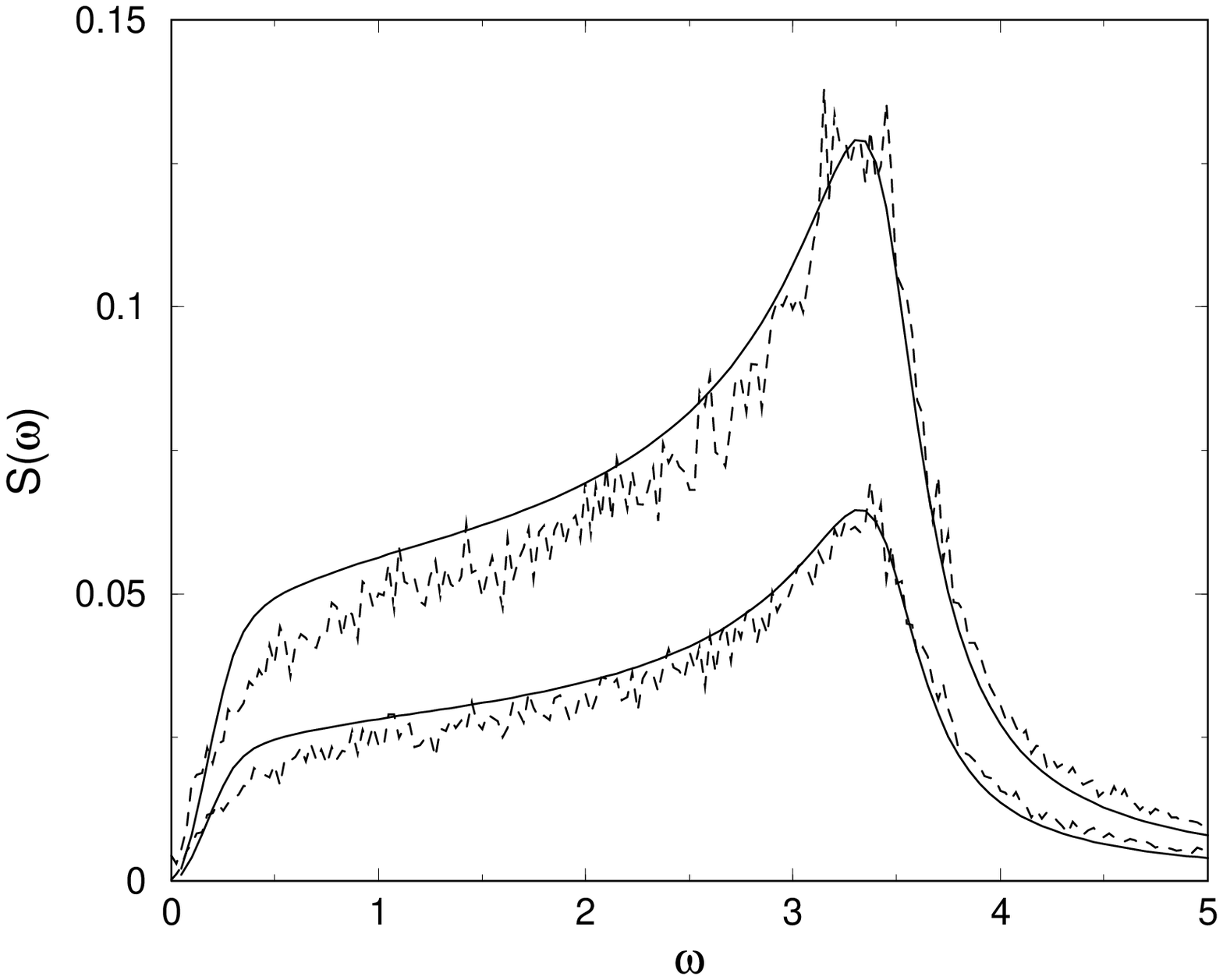}
\end{center}
\caption
{Spectrum $S_n^0(\omega)$ for any site at two temperatures, $k_BT=0.05$ (lower
curves) and $k_BT=0.1$ (upper curves).  The smooth curves are the
predictions in Eq.~(\ref{predb}) and the jagged curves are the numerical
simulation results averaged over 500 runs.  The irregularities can
be eliminated at great CPU time cost by refining the time and/or frequency
meshes and/or increasing the number of runs.  Other parameters: $k=3$,
$\gamma=0.5$.
}
\label{appfig1}
\end{figure}

Figure \ref{appfig1} shows the predicted spectrum (smooth curves) and
the corresponding simulation results for two different temperatures.
The agreement is typical of broad parameter ranges.

\subsection{Finite Temperature and Signal: SNR}

Finally, we deal with the full equation of motion
\begin{equation}
\ddot x_n   =   k(x_{n+1} +x_{n-1} - 2x_n) - \gamma \dot{x}_n + f_n(t)
\label{clang}
\end{equation}
with the signal $\dot{x}_0  = A  \sin ( \omega_0 t)$ applied at site
$n=0$.  The
fluctuation $f_0$ is set to zero since the velocity of that site
is fixed by the signal.

The Fourier transform is defined as before, but in transforming the
equations of motion we must take into account that (\ref{clang}) does not
hold for $n=0$.  We thus multiply (\ref{clang}) by $e^{2\pi iqn}$ and sum
over $n$ but only from $n=1$ to $N-1$:
\begin{equation}
\sum_{n=1}^{N-1} x_n e^{2\pi i qn} = y_q-x_0.
\end{equation}
Proceeding in this manner we obtain in place of Eq.~(\ref{ftheequation})
\begin{equation}
\ddot{y}_q + 4 k \sin^2\left( \frac{\pi q}{N} \right) y_q + \gamma
\dot{y}_q \ = \ U_q(t)
\label{fntheequation}
\end{equation}
where the inhomogeneous term now is
\begin{equation}
U_q(t) \ = \ F_q(t) + k\left( 2x_0 - x_1 - x_{N-1}\right) + \gamma
\dot{x}_0 + \ddot{x}_0.
\end{equation}
The terms containing $x_0$ and its derivatives are known; $x_1$ and
$x_{N-1}$ (they are equal by symmetry) would have to be found
``self-consistently" ex-post (see below).

Equation~(\ref{fntheequation}) is again a linear inhomogeneous differential
equation with constant coefficients.  Its solution has two additive
contributions:
\begin{equation}
y_q(t) \ = \ y_q^{hom}(t) + y_q^{inh}(t)
\end{equation}
The portion $y_q^{hom}(t)$ is the solution of the homogeneous
equation, that is, with $U_q(t)=0$, and the contribution $y_q^{inh}(t)$
is due to the inhomogeneity.

Suppose that $F_q(t)$ were equal to zero.  The solution in this case,
which we denote by $y_q^s(t)$ is exactly the solution of Sec.~\ref{appa};
that is, the inverse transform of
\begin{equation}
y_q^s(t) \ = \ y_q^{s,hom}(t) + y_q^{s,inh}(t)
\label{previous}
\end{equation}
at long times must be precisely Eq.~(\ref{solution}).
Now consider the consequences of again including $F_q(t)$.
The homogeneous part of the solution has to be exactly as
before, i.e., $y_q^{hom}=y_q^{s,hom}$ because the homogeneous part of
the equation has not changed, the constraint at $x_0$ also has not
changed, and initial conditions are in any case immaterial at long times.
However, the inhomogeneous part of the solution does
change, and it does so in two ways.  First and most important is the
direct contribution of the thermal fluctuations to the solution.  Second,
and less important, is the {\em change} in $x_1$ and $x_{-1}$ (which is
the same as that of $x_1$) caused by the addition of the noise, and
the change that this in turn imposes on the other displacements.  This
contribution is more difficult to calculate and affects only sites near
$n=0$.  Ignoring the latter we thus assume the solution
\begin{equation}
y_q(t) \ = \ y_q^s(t) + y_q^0(t)
\end{equation}
where $y_q^s(t)$ is given in Eq.~(\ref{previous}) and $y_q^0(t)$ in
Eq.~(\ref{thisone}).

\begin{figure}[htb]
\begin{center}
\leavevmode
\epsfxsize = 3.0in
\epsffile{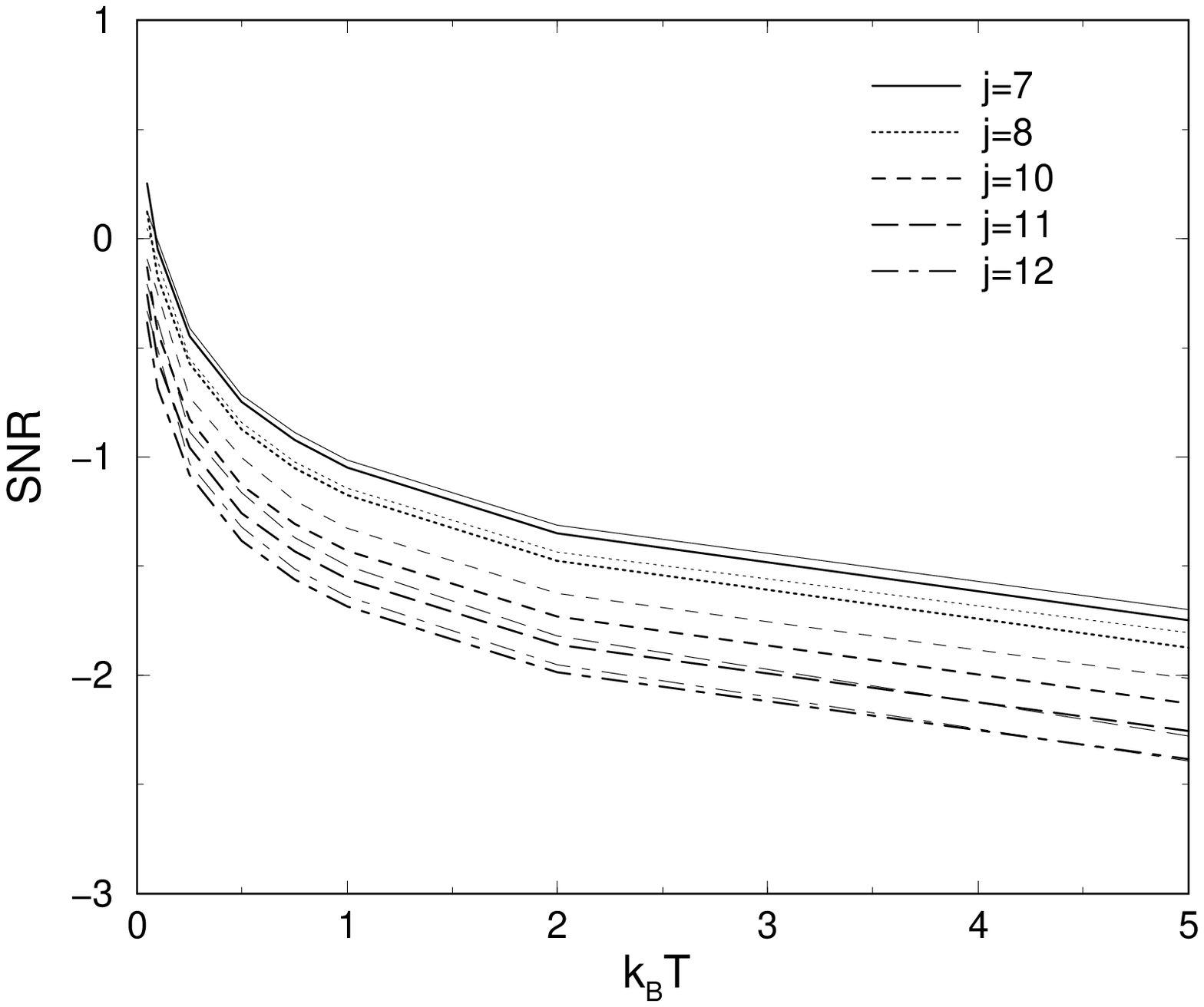}
\leavevmode
\hspace{0.2in}
\epsfxsize = 3.0in
\epsffile{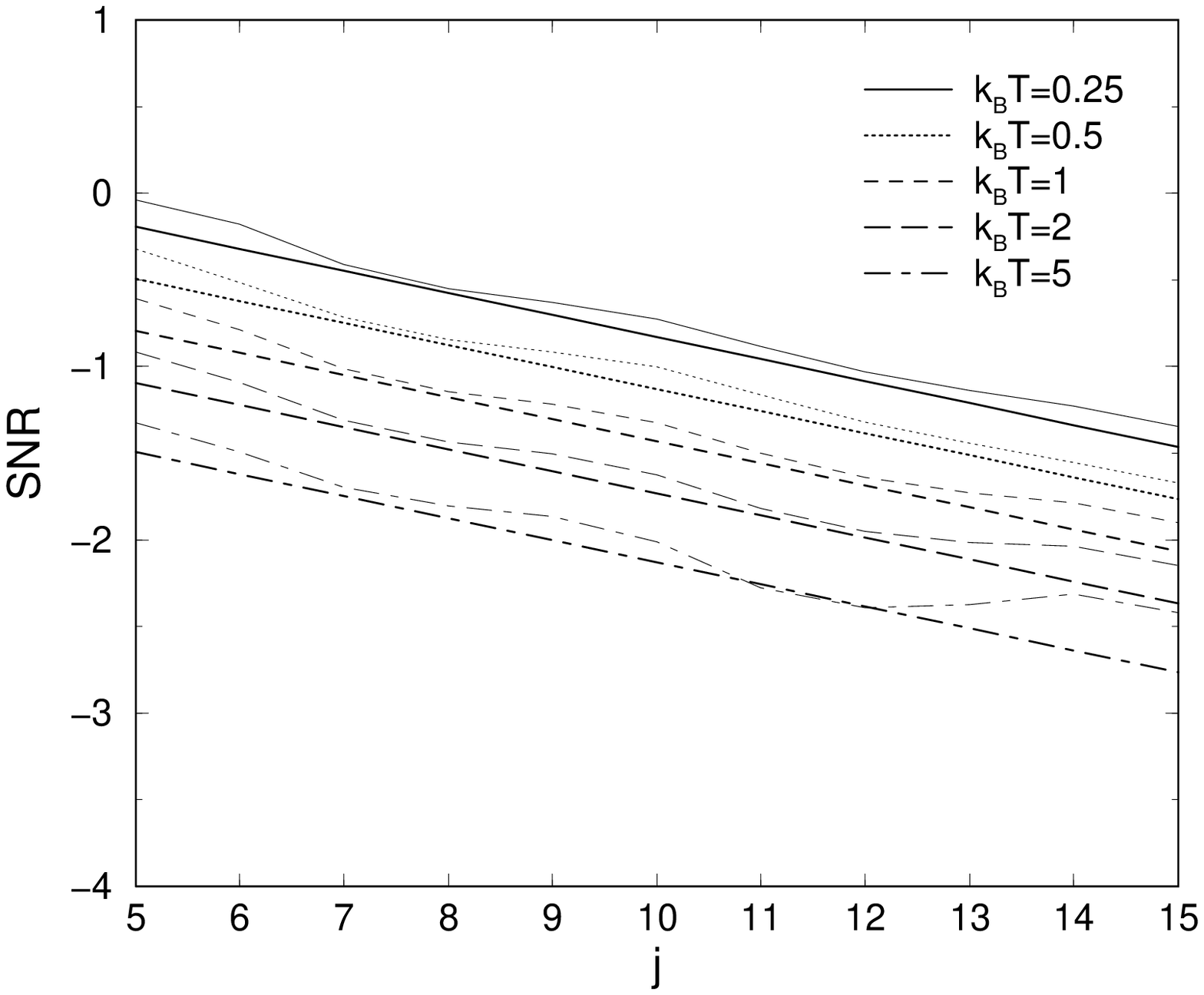}
\end{center}
\caption
{First panel: Typical $SNR$ curves as a function of temperature
for sites 7, 8, 10, 11, and 12 along a harmonic
chain with $k=3$, $\gamma=0.5$, $\omega_0=1.0$, and $A=0.5$.
The numerical simulation results (thin lines) in each case lie above the
analytic curves (thick lines).
Second panel: $SNR$ as a function of site for different temperatures.}
\label{comparisons}
\end{figure}

The correlation function and spectrum are then simply the sums of the
correlation functions and spectra for the system with the signal but no
fluctuations (zero temperature) and those for the thermalized system in the
absence of the signal.  Thus, the spectrum is the sum of (\ref{asignal})
(appropriately weighted by a delta function that places it at $\omega_0$)
and (\ref{predb}).  In particular, the SNR for the harmonic system then
finally is
\newcommand{\D}{\displaystyle}
\begin{equation}
SNR _n(\omega_0) = \log_{10}
\left(
\ \frac{ \frac{\D A^2}{\D 2} e^{-2n\mu}}
{\frac{ \D 2\gamma k_BT\omega^2}{\D N}\sum_{q=0}^{N-1}
\frac{\D 1} {\D [r_1 ^2(q)+\omega^2] [r_2 ^2(q)+\omega^2]}} \
\right)
\label{thatsit}
\end{equation}
which decreases monotonically with $T$ as well as with $n$.
This result is consistent with the linear response theory approach
introduced and widely applied by Dykman et al.~\cite{LRT}.
Comparisons of this analytic result with numerical simulations
are shown in Fig.~\ref{comparisons} (cf. Fig.~\ref{figsnrosch}).
Equation~(\ref{thatsit}) clearly captures the correct behavior;
the small discrepancies between analytic and numerical results
(which always lie above the analytic curves) reflect the omitted
terms discussed above.

\end{document}